\documentclass[preprint,12pt]{aastex}
\usepackage{natbib,amsmath}
\newcommand {\apgt} {\ {\raise-.5ex\hbox{$\buildrel>\over\sim$}}\ }
\newcommand {\aplt} {\ {\raise-.5ex\hbox{$\buildrel<\over\sim$}}\ } 
\shorttitle{The Column Temperature}
\shortauthors{Shetty et al.}

\begin{document}
\title{The Effect of Line of Sight Temperature Variation and Noise on Dust Continuum Observations}
\author{Rahul Shetty\altaffilmark{1,2}, Jens Kauffmann\altaffilmark{1,2}, Scott Schnee\altaffilmark{3}, Alyssa A. Goodman\altaffilmark{1,2}, Barbara Ercolano\altaffilmark{1,4}}
\altaffiltext{1}{Harvard-Smithsonian Center for Astrophysics, 60
  Garden Street, Cambridge, MA 02138}
\altaffiltext{2}{Initiative for Innovative Computing, Harvard University, 60 Oxford Street, Cambridge, MA, 02138}
\altaffiltext{3}{Division of Physics, Mathematics, and Astronomy, California Institute of Technology, 770 South Wilson Avenue, Pasadena CA 91125}
\altaffiltext{4}{Institute for Astronomy, Madingley Rd, Cambridge, CB3 OHA, UK}
\email{rshetty@cfa.harvard.edu}

\begin{abstract}

We investigate the effect of line of sight temperature variations and
noise on two commonly used methods to determine dust properties from
dust continuum observations of dense cores.  One method employs a
direct fit to a modified blackbody SED; the other involves a
comparison of flux ratios to an analytical prediction.  Fitting fluxes
near the SED peak produces inaccurate temperature and dust spectral
index estimates due to the line of sight temperature (and density)
variations.  Longer wavelength fluxes in the Rayleigh-Jeans part of
the spectrum (\apgt 600 \micron\ for typical cores) may more
accurately recover the spectral index, but both methods are very
sensitive to noise.  The temperature estimate approaches the density
weighted temperature, or ``column temperature,'' of the source as
short wavelength fluxes are excluded.  An inverse temperature -
spectral index correlation naturally results from SED fitting, due to
the inaccurate isothermal assumption, as well as noise uncertainties.
We show that above some ``threshold'' temperature, the temperatures
estimated through the flux ratio method can be highly inaccurate.  In
general, observations with widely separated wavelengths, and including
shorter wavelengths, result in higher threshold temperatures; such
observations thus allow for more accurate temperature estimates of
sources with temperatures less than the threshold temperature.  When
only three fluxes are available, a constrained fit, where the spectral
index is fixed, produces less scatter in the temperature estimate when
compared to the estimate from the flux ratio method.
\end{abstract}


\section{Introduction\label{introsec}}

Some of the coolest regions in molecular clouds are dense, starless
cores.  These dust enshrouded objects are often in the process of
forming one (or a few) protostar(s) \citep{Benson&Myers89}.
Determining the physical properties of cores, such as temperature,
composition, and density, is necessary for a complete understanding of
the environmental conditions prior to the formation of a star or
protostar.  There has been much progress in the study of cores
containing central protostars, including the success of theory in
explaining the variety of emergent spectral energy distributions (SED)
as an evolutionary sequence
\citep[e.g.][]{Adamsetal87,Lada87,Andreetal93}.  On the other hand,
the structure and evolution of cores that have yet to form a central
protostar is not as well understood, and is thus an active area in
current star formation research.

Dust presents one avenue to observationally investigate starless
cores.  Dust is prevalent in the ISM, and it is responsible for most
of the extinction of starlight.  For cores positioned in front of
sources of known luminosity or color, the level of extinction can be
an indicator of the dust content in the attenuating core
\citep{Ladaetal94,Alvesetal01}.  Additionally, scattered light from
cores surrounded by diffuse background radiation may be used to
determine the dust content \citep{Foster&Goodman06}.  Dust can also be
directly detected through its thermal emission.  Since the
temperatures of the cores are $\aplt$ 15 K, the emergent continuum SED
peaks in the far-infrared (FIR) or sub-millimeter wavelength regimes
(see Fig. \ref{sedintro}).  Ground and space based observations by
{\it SCUBA}, {\it MAMBO}, {\it Bolocam}, {\it 2MASS}, {\it IRAS}, {\it
ISO} and {\it Spitzer} have detected dust emission from many
environments, and they have provided much information about starless
cores \citep[e.g.][]{WardThompsonetal02,Schneeetal07,Kauffmannetal08}.
The upcoming {\it Planck} and {\it Herschel} missions, which are
capable of FIR observations, are also well suited for detecting dust
emission.  Thus, a thorough consideration of the nature of dust
continuum emission, and the uncertainty associated with measuring it,
is timely.

The main characteristics of the dust that determine the form of the
emergent SED are the column density, temperature, and emissivity.
Observationally quantifying these characteristics should constrain
models of dense starless cores.  Radial density profiles are often
compared with a stable isothermal Bonnor-Ebert sphere
\citep{Bonnor56,Ebert55}, for which the volume density is constant
near the center, but drops as $r^{-2}$ at larger radii
\citep[e.g.][]{Bacmannetal00,SchneeGoodman05}.  The density gradients
may vary from core to core, which in turn may (or may not) be due to
an evolutionary sequence as cores continually collapse to form a
protostar.

Though gas temperatures in a stable Bonnor-Ebert sphere are constant,
theoretical results have suggested that dust temperatures decrease
towards the center to values as low as $\sim$ 7 K
\citep[e.g.][]{Leung75,Evansetal01,Zucconietal01}.  And, recent
observational investigations have in fact identified cores with such
gradients in the dust temperatures
\citep[e.g.][]{Schneeetal07,WardThompsonetal02}.  Though the dust mass
is only a fraction of the gas mass ($\sim$ 1/100), gas temperatures
may also exhibit gradients, due to the coupling between dust and gas
at high enough densities \citep{Goldsmith01,Crapsietal07}.  The
(recent and upcoming) availability of higher quality observational
data will require a thorough interpretation of emergent SEDs to
accurately assess the temperature, as well as density, profiles of the
observed sources.

The common assumption is that the emergent SED from interstellar dust
is similar to the Planck function of a blackbody, modified by a
power-law dependence on the frequency \citep{Hildebrand83}.  The
spectral index of the dust emissivity power-law, $\beta$, is dependent
on the bulk and surface properties of the dust grains.  As shown by
\citet{Keeneetal80}, observations limited by sparse flux sampling may
be consistent with various SEDs described by different values of
$\beta$.  A precise estimate of the value of $\beta$ is necessary to
accurately derive other properties of the observed source, such as the
temperature and the mass of a cold core.  The emissivity-modified
blackbody spectrum is the basis for many analyses of dust properties
in observed cores
\citep[e.g][]{Krameretal03,Schneeetal05,WardThompsonetal02,Kirketal07}.

Dupac and coworkers fit observed FIR and sub-millimeter fluxes with a
modified blackbody spectrum, and they suggested that $\beta$ decreases
with increasing temperatures, from $\sim$ 2 in cold regions to 0.8 -
1.6 in warmer regions \citep[$T \sim$ 35 - 80
K;][]{Dupacetal01,Dupacetal02,Dupacetal03}.  Such analyses may be
sensitive to the simplified assumption of a constant temperature along
the line of sight.  Using radiative transfer calculations of embedded
sources, \citet{Doty&Leung94} demonstrated that the accuracy of the
parameters estimated from measured fluxes is sensitive to the precise
nature of the source (e.g. opacity, temperature distribution); they
found that the Rayleigh-Jeans (R-J) regime of the emergent spectrum is
better suited for an accurate determination of the dust spectral
index.  \citet{Doty&Palotti02} found that the use of flux ratios to
estimate the spectral index is sensitive to which wavelengths (of the
given fluxes) are used in the ratio; they also found that $\beta$ is
more accurately determined when fluxes at longer wavelengths are used
in a fit.  \citet{Schneeetal06} also showed that various ratios of
fluxes (with different wavelengths) give different estimates for dust
temperature and column density, due to an inaccurate isothermal
assumption.

Here, we systematically investigate how line of sight density and
temperature variations, similar to those in dense cores, as well as
noise uncertainties, affect the temperature and spectral index
estimated from IR and sub-millimeter observations.  We focus on two
commonly employed methods.  The first method uses a direct fit of a
modified blackbody SED.  For the second method, ratios of the fluxes
are used to determine the temperature and $\beta$ from an analytical
prediction.  Both methods usually rely upon an assumption of constant
temperature along the line of sight.  Using simple radiative transfer
calculations of model sources, and Monte Carlo experiments, we assess
how well the resulting temperature and spectral index estimates
recover properties of known input sources.

This paper is organized as follows.  In the next section ($\S$2), we
present the analytical expression of the power-law modified blackbody
spectrum, and briefly introduce two well known methods used to
estimate the dust properties from IR and sub-millimeter continuum
observations.  In our analysis of the two methods, we consider
numerous scenarios typical of observations of star forming regions.
Table \ref{overv} shows the particular scenario considered in each
subsection, and may be used as a brief guide to $\S$3 - $\S$6.  We
begin our analysis by considering non-isothermal sources in the ideal
limit where a large range of fluxes at different wavelengths are
available for fitting an SED.  We then systematically exclude fluxes,
culminating with the scenario where only a few fluxes are available,
in which case the flux ratio method is employed.  Throughout our
analysis, we also consider the effect of noise in the observations, of
both isothermal and non-isothermal sources.  In $\S$\ref{fitsec}, we
describe the method to estimate source temperatures using direct SED
fitting; we investigate how line of sight variations, noise, and the
sampling of different regions of the emergent SED affect the resulting
temperature estimates.  We also discuss our findings in the context of
recent published works.  In the following section ($\S$\ref{flxsec})
we analyze the flux ratio method focusing on the effect of noise,
through the use Monte Carlo simulations.  We then compare the two
methods using a radiative transfer simulation to model the emission
from an isolated starless core in $\S$\ref{comparesec}.  After a
discussion in $\S$\ref{discsec}, we summarize our findings in
$\S$\ref{sumsec}.

\section{Dust Emission: Common Assumptions and Methods \label{methodsec}}
\subsection{Isothermal Sources}

The emergent continuum SED due to dust is often expressed analytically
as the product of a blackbody spectrum $B_\nu(T)$ at the dust
temperature $T$ and the frequency dependent dust opacity $\kappa_\nu$.
The observed flux density associated with this SED takes the form
\begin{equation}
S_\nu = \Omega B_\nu(T)\kappa_\nu N,
\label{fd}
\end{equation}
where $\Omega$ is the solid angle of the observing beam, and $N$ is
the column density of the emitting material.  The opacity $\kappa_\nu$
is empirically determined to have a power law dependence on the
frequency \citep{Hildebrand83}:

\begin{equation}
\kappa_\nu = \kappa_0 \left( \frac{\nu}{\nu_0} \right)^\beta.
\label{opac}
\end{equation}
The spectral index $\beta$ depends on the physical and chemical
properties of the dust.  For silicate and graphite dust composition
common in much of the ISM, $\beta \sim 2$ \citep{Draine&Lee84}.
However, observations have shown that $\beta$ can reach values as low
as \aplt 1 and as high as \apgt 3 in various environments
\citep[e.g.][]{Oldhametal94,Kuanetal96,Mathis90}.  Indeed, the
spectral index is a key parameter and accurately determining its
value, along with the column density $N$ and the temperature $T$, is
crucial for a thorough description of dust properties in an observed
region.  Those are the three parameters that are required to
accurately describe an observed flux density (per beam,
i.e. $S_\nu/\Omega$).

Figure \ref{sedintro} shows SEDs from a 20 K source with different
values of $\beta$, but constant column densities.  For comparison, the
SED from a 5 K source with the same column density and $\beta$ = 2 is
also shown. The SEDs are all calibrated using an equivalent $\kappa_0$
= $\kappa_{230 \, {\rm GHz}}$, which is why all the 20 K SEDs
intersect at 230 GHz.  Sources with higher spectral indices ($\beta$)
produce SEDs with steeper slopes at long wavelengths (in the R-J
regime), increasing peak fluxes, and shorter peak wavelengths.
Similar to Wien's Law for a pure blackbody, for a given value of
$\beta$, a modified Wien's Law indicating the wavelength corresponding
to the peak in $S_{\nu}$, $\lambda_{max}$, can be determined
numerically.\footnote{This modified Wien's Law gives the wavelength
that corresponds to the peak in $S_{\nu}$.  In other texts,
$\lambda_{max}$ sometimes refers to the wavelength corresponding to
the peak of $S_{\lambda}$.}  For $\beta$ = 2, $\lambda_{max} \simeq
(2900\, \mu {\rm m}\, K)/T$, and for $\beta$=1, $\lambda_{max} \simeq
(3670\, \mu {\rm m}\, K)/T$.  \citet{Doty&Palotti02} find that
$\lambda_{max} = (4620 e^{-0.2357 \beta} \, \mu {\rm m}\, K)/T$ is a
good fit for $1 < \beta < 2$.

\subsection{Non-Isothermal Sources: A Simple Example}

Equations (\ref{fd})-(\ref{opac}) describe the spectrum emitted from
dust at a single temperature $T$.  For a 3D source with various dust
characteristics, the emergent SED will be a combination of SEDs from
all the dust in the source.  For optically thin emission, the emergent
SED is simply the integrated SED from each dust grain.  Here, we
briefly consider the effect of using equations (\ref{fd})-(\ref{opac})
to characterize an SED from a source with two different dust
populations.  This relatively simple analysis is a prelude to the
effect of line of sight temperature variations on commonly employed
methods to estimate dust properties.

Figure \ref{2compfigsed} shows the emergent SED from a source with two
populations of dust grains, along with the SED from each individual
component: the temperature and column density of the cool component is
$T_1$ = 10 K and $N_1$, respectively; for the warm component, $T_2$ =
15 K and $N_2 = 0.1 N_1$.  Physically, such a system is similar to a
10 K dense core surrounded by a 15 diffuse envelope.  The spectral
indices for both components are set to $\beta$ = 2.

The peak of the emergent SED in Figure \ref{2compfigsed} occurs at
$\lambda_{max}$ = 251 \micron.  Using the modified Wien's Law for
$\beta$=2, the temperature of an isothermal source that would produce
a spectrum which peaks at that wavelength is 11.6 K.  Though this
temperature occurs somewhere between the temperatures of the two
isothermal sources that contribute to the emergent SED, in practice
knowledge of the peak of the SED of an unknown source is not easily
determined.  Further, it is not obvious how one should interpret the
temperature assigned to a source which itself is not isothermal.

In subsequent sections, we assess how well emergent SEDs are described
by equations (\ref{fd})-(\ref{opac}), and what information about the
source temperature can be garnered from continuum observations that
span different regions of the SED.  We also consider the more
realistic constraint of limited sampling of the emergent SED.  One
question we aim to address, for instance, is whether fluxes in
different parts of an emergent SED, such as the Wien or R-J regimes,
are preferable for determining the dust properties.

Two commonly employed methods to determine the properties of dust from
continuum observations are: (1) a direct fitting of equations
(\ref{fd}) \& (\ref{opac}); and (2) the use of ratios of observed flux
densities at 2 or more wavelengths.  For isothermal sources, and with
ideal observations with no uncertainties, both methods will accurately
recover $T$, $\beta$, $N$, and $\kappa_\nu$.  However, all sources are
unlikely to be isothermal, and even the most accurate observations
include some level of intrinsic noise.  In the following sections, we
quantify how these factors affect the accuracy of the derived
parameters.  After describing the two methods in further detail, we
use simple numerical experiments to evaluate the accuracy of the
methods in determining the dust properties from observations of star
forming cores.

\section{Direct SED Fitting \label{fitsec}}

A minimized $\chi^2$ fit of equations (\ref{fd})-(\ref{opac}) to a
number of observed fluxes can be performed to estimate the dust
properties.  There are essentially three parameters to be fit, the
temperature $T$, the spectral index $\beta$ and the absolute scaling,
which is just the product of the column density $N$ and the opacity at
a given frequency $\kappa_0$.  Since a fit will only produce the
scaling (which is the optical depth at a particular frequency,
e.g. $\tau_{230} = N\kappa_{230}$ at 230 GHz), other assumptions
and/or techniques are necessary to obtain estimates of $N$ and
$\kappa_0$.  For example, if the opacity at a wavelength is known
(e.g. $\kappa_0=\kappa_{230}$ for $\nu_0=$ 230 GHz), then the fit can
estimate $N$ directly.  Extinction studies are another avenue to
estimate $N$; the level of attenuation (usually from optical and NIR
observations) due to dust in dark clouds in front of the stellar
background is directly related to the column density of the dust
\citep[e.g.][]{Ladaetal94}.  This method is advantageous since it
provides an independent estimate of $N$, but also requires
assumptions, such as the ratio of total-to-selective extinction $R_V$
\citep[e.g.][and references therein]{Hildebrand83,Mathis90}.  Further,
to obtain the total column density along the line of sight, and not
just that of the dust, an additional assumption of the dust-to-gas
ratio is required.  In our analysis, we will assume that only IR and
sub-millimeter observations are available, and thus will limit our
analysis to the estimation of $T$ and $\beta$.  Our focus here is to
investigate how well a given method can reproduce temperatures and
spectral indices only; estimation of the absolute column density and
opacity is beyond the scope of this work.

\subsection{Effect of Line of Sight Temperature Variations}

\subsubsection{Two-Component Sources \label{2comsec}}

We begin by considering ideal (i.e. error-free) observations of simple
two-component sources.  Fitting experiments involving sources with two
dust populations have been explored by \citet{Dupacetal02}; their aim
was to determine the amount of cold dust, along lines of sight with
warmer dust, that is necessary to reproduce the fit results of their
observations.  Here, we simply evaluate the resulting fits when fluxes
at different wavelengths are available.

The temperatures of the cold and warm media are $T_1$ and $T_2$,
respectively, and the column density ratio is $N_2/N_1$.  Such systems
are analogous to isothermal (spherical) dense cores surrounded by
warmer envelopes.  Modified blackbody SEDs
(eqns. [\ref{fd}]-[\ref{opac}]) are constructed for the two media,
both with $\beta$ = 2.  A particular example SED of this general case
is shown in Figure \ref{2compfigsed}.  We then fit equation (\ref{fd})
to the integrated SEDs, solving for $T$ and $\beta$ (as well as the
scaling factor $N \kappa_0$).  Figure \ref{2compfig} shows the fit
temperatures and spectral indices from observations of a variety of
two-component systems, along with the wavelength range of the fluxes
considered in the fit.

Since $T$ varies along the line of sight, the best fit $T$ will likely
not be equal to the temperature of one of the two sources.  One
characteristic temperature of this two-component medium is the density
weighted temperature \citep[e.g.][]{Doty&Palotti02}.  Since this
density weighted temperature is analogous to the column density, we
will call it the ``column temperature,'' $T_{col}$.  The estimated
temperature from a fit can be compared with this true column
temperature.

As shown in Figure \ref{2compfig}, the best fit temperature is
systematically too high when all fluxes at (integer) wavelengths
between 10 - 3000 \micron\ are considered in the fit.  In fact, when
the temperature difference between the two components is large, the
best fit temperature is actually larger than the warmer medium, as in
``2COMPc'' and ``2COMPd,'' lines of sight containing dust at 10 and 20
K.  Further, when all wavelengths are considered, the fit value of
$\beta$ is always lower than the actual value of 2.  However, when
shorter wavelengths are systematically excluded from the fit, the best
fit temperature decreases and approaches the column temperature.  The
best fit $\beta$ also approaches the model value of 2 when fluxes in
the R-J wavelength regime are the only ones used in the fit,
consistent with the findings of \citet{Doty&Leung94}.  In practice,
for deriving dust properties from the emergent SED, it may be
necessary to exclude fluxes with $\lambda$ \aplt\ 100 \micron\ due to
the contribution of embedded sources as well as transiently heated
very small grains \citep{Li&Draine01}, depending on the environment.

\subsubsection{Cores with Density and Temperature Gradients \label{corevarT}}

Recent theoretical and observational studies have indicated that the
dust temperature in starless cores decreases toward the center,
reaching low values \aplt 7 K
\citep[e.g.][]{Evansetal01,Crapsietal07,Schneeetal07,WardThompsonetal02}.
We thus investigate the emergent SED from cores containing temperature
gradients like those observed, and whether any useful information can
be obtained from fitting a single power-law modified blackbody
spectrum to that SED.

To construct the model cores, we use the density and dust temperature
profiles presented by \citet{Evansetal01}, who performed radiative
transfer simulations on a variety of model cores with a range of
density profiles.  Though the resulting dust temperature profiles
$T(r)$ are sensitive to the model density profiles, the relationship
between $T$ and column density $N$ is relatively uniform between
models with different (volume) density profiles \citep[see Fig. 9
of][]{Evansetal01}.  In our analysis of emergent SEDs from starless
cores, we construct two cores, with temperatures ranging between 8 -
12 K and 5 - 12 K; the column densities are as indicated in Figure 9
of \citet{Evansetal01}: $N$ increases with decreasing temperature (and
thus with decreasing core radius).  At the outer edge of the core, at
a temperature of 12 K, the column density is set to 2$\times 10^{21}$
cm$^{-2}$; the temperature drops to 8 K in model Core 1 and to 5 K in
model Core 2, with column densities of 1.25$\times 10^{22}$ cm$^{-2}$
and 1$\times 10^{23}$ cm$^{-2}$, respectively.  Besides these isolated
cores, without any surrounding medium, we also consider cases where
the cores are surrounded by an envelope with a temperature of 20 K and
a column density of 1$\times 10^{21}$ cm$^{-2}$.

As described in $\S$\ref{2comsec} we begin by fitting the emergent SED
assuming that fluxes at various wavelength ranges are
available. Figure \ref{modcores} shows the resulting best fit
temperatures and spectral indices for the two cores.  The SEDs from
the cores without an envelope are analogous to an SED obtained by
accurately subtracting off flux due to larger scale emission from the
surrounding region, or an SED from a truly isolated core.  When
equation (\ref{fd}) is fit to fluxes at 100 - 600 \micron, the best
fit temperature is $\sim$ 3 K ($\sim$ 15\%) off from the column
temperature of Core 1, but differs from the column temperature of Core
2 by $\sim$ 6 K ($\sim$ 50\%).  The best fit spectral index also shows
large variation between the two cores.  For Core 1 ($T \in 8-12$ K)
the fit $\beta$ of 1.65 is within 20\% of the model value of 2.
However, for Core 2 ($T \in 5-12$ K), the best fit $\beta$ of 0.81 is
erroneous by over a factor of 2.  As more short wavelength fluxes (in
the Wien regime) are excluded, the fits recover the model spectral
index more accurately; the temperature estimate also decreases,
approaching the column temperature of the core.

When the core is surrounded by a warmer envelope, or when the flux
from extended regions has not been properly accounted for, then the
discrepancy between the best fit parameters and the core properties
increases, as expected.  For Core 2, a fit to the fluxes at 100 - 600
\micron\ results in an estimate for $\beta$ with an unphysical sign
(-0.3).  Including the envelope, the discrepancy between the fit $T$
and the column temperature at short wavelength fluxes increases by
$\sim$50\%, compared with the cores without the envelope.

We performed such fits for cores surrounded by more diffuse envelopes,
with a column density that is a factor $\sim$ 10 - 100 times lower
than that of the core (1$\times 10^{20}$ cm$^{-2}$).  Such envelopes
only have a slight effect on the fit temperatures, because the column
temperatures are not significantly different compared to the isolated
cores.  The envelope does have an appreciable effect on the best fit
$\beta$ when fluxes near the peak of the SED are considered in the
fit.  However, excluding short wavelength fluxes still recovers the
true spectral index reasonably accurately, as Figure \ref{sedintro}
would suggest.

Figure \ref{coresed} shows the emergent SED from Core 2 ($T \in$ 5-12
K) without an envelope, along with the results from two fits, one to
the fluxes from 100 - 600 \micron, and the other to fluxes from 1 - 3
mm.  The fit at shorter wavelengths does indeed reproduce the peak and
shorter wavelength fluxes of the emergent SED reasonably well, but
severely overestimates the fluxes at longer wavelengths.  On the other
hand, the fit to long wavelength observation reproduces the R-J tail
of the spectrum accurately, but underestimates the fluxes at all
shorter wavelengths.

The peak of the emergent SED from Core 2 occurs at $\lambda$ = 324
\micron.  From the modified Wien's Law (for $\beta$=2) $\lambda_{max}
= (2900\, \mu {\rm m}\, K)/T$, the temperature associated with this
wavelength is 8.9 K.  This value is closer to the maximum temperature,
12 K, of the source, than the column temperature $T_{col}$=6.2 K.  In
this case, $T_{col}$ is dominated by the high density, cold center,
whereas the peak of the SED is primarily influenced by the warmer, low
density regions of the core.  Since the SED has an exponential
dependence on the temperature in the Wien regime, which, for such a
cold core, goes up to $\sim$ 100 \micron, even low density regions may
dominate the total SED emerging from a region, due to the higher
temperatures.

\subsubsection{Summary of the Effect of Line of Sight Temperature Variations} 

As expected, the emergent SED at wavelengths near the peak is poorly
described by a simple power-law-modified-blackbody, due to temperature
variations along the line of sight, as previously documented in the
literature \citep[e.g.][]{Doty&Palotti02,Schneeetal06}.  Nevertheless,
the fits can still reveal properties of the observed source, namely
that the best fit temperature approaches the column temperature and
the best fit $\beta$ approaches the model value as shorter wavelengths
are excluded.  There are also other revealing trends that warrant
further investigation.  First, the systematic exclusion of short
wavelength fluxes results in different $T$ and $\beta$ estimates; for
an isothermal source, the fit $T$ would always be the same (and equal
to the temperature of the source).  If this trend also occurs when
there are only a few observations, then short wavelength observations
can still be used to determine whether a source is isothermal or not.
Second, there also appears to be an inverse $T$ - $\beta$
relationship: whenever $T$ is overestimated, $\beta$ is
underestimated.  A similar relationship has been discussed by Dupac
and coworkers using observations at wavelengths $<$ 600 \micron\
\citep{Dupacetal01,Dupacetal02,Dupacetal03}.  A thorough analysis of
the sources observed by Dupac et al. would be warranted to rule out
that the inferred anti-correlation is simply due to line of sight
effects.  As we discuss in the next section ($\S$\ref{isonoise}), an
inverse $T$ - $\beta$ trend may also arise from SED fits due solely to
noise in the observations.

\subsection{Effect of Noise \label{isonoise}}

Uncertainties in observed fluxes may also lead to incorrect
temperature and spectral index estimates from SED fitting.  To assess
the effect of noise, we first consider fluxes from isothermal sources
with modest 5\% uncertainties in each observed flux.  A number fluxes
from a range of (integer) wavelengths are considered in the fit: 100 -
600 \micron, 500 - 1000 \micron, and 1000 - 1500 \micron.  In these
Monte-Carlo experiments, each flux is modified by a random value drawn
from a Gaussian distribution, with $\sigma$ = 0.05.

Figure \ref{noisefig} shows the best fit $T$ and $\beta$ estimates for
isothermal sources, one with $T$ = 10 K and $\beta$=2, and the other
with $T$ = 20 K and $\beta$=2.  Each set of noisy fluxes, spanning the
different wavelength regimes, is generated 100 times; a modified
blackbody is then fit to each set.  As expected, there is a spread in
the best fit $T$ and $\beta$.

\begin{figure}
\plotone{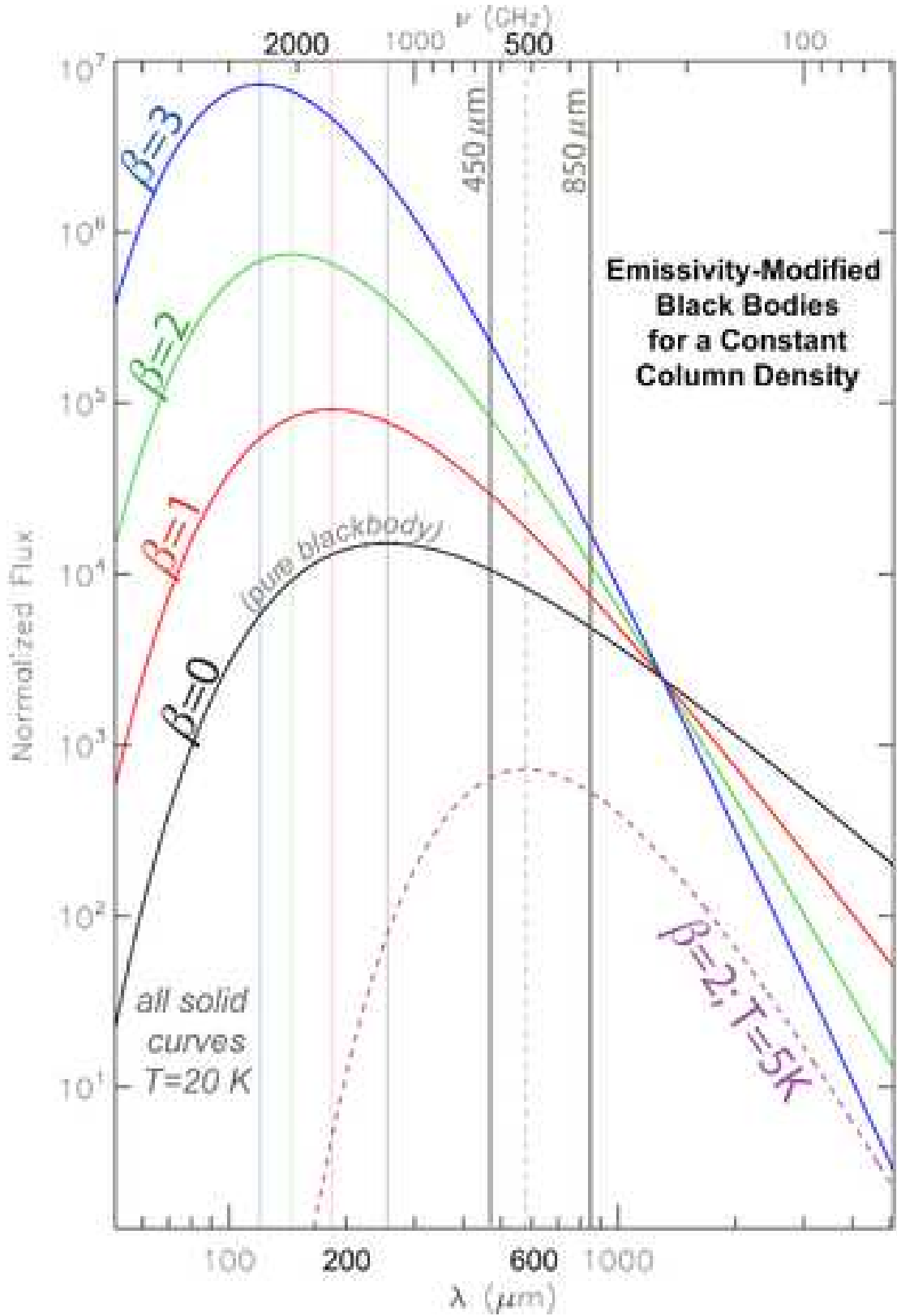}
\caption{Emissivity modified blackbodies with different spectral
indices $\beta$, but constant column density, from a 20 K source.
Dashed SED is from a 5 K source with $\beta$ = 2.  Thin solid vertical
lines indicate the peak wavelength of the SEDs.}
\label{sedintro}
\end{figure}

\begin{figure}
\plotone{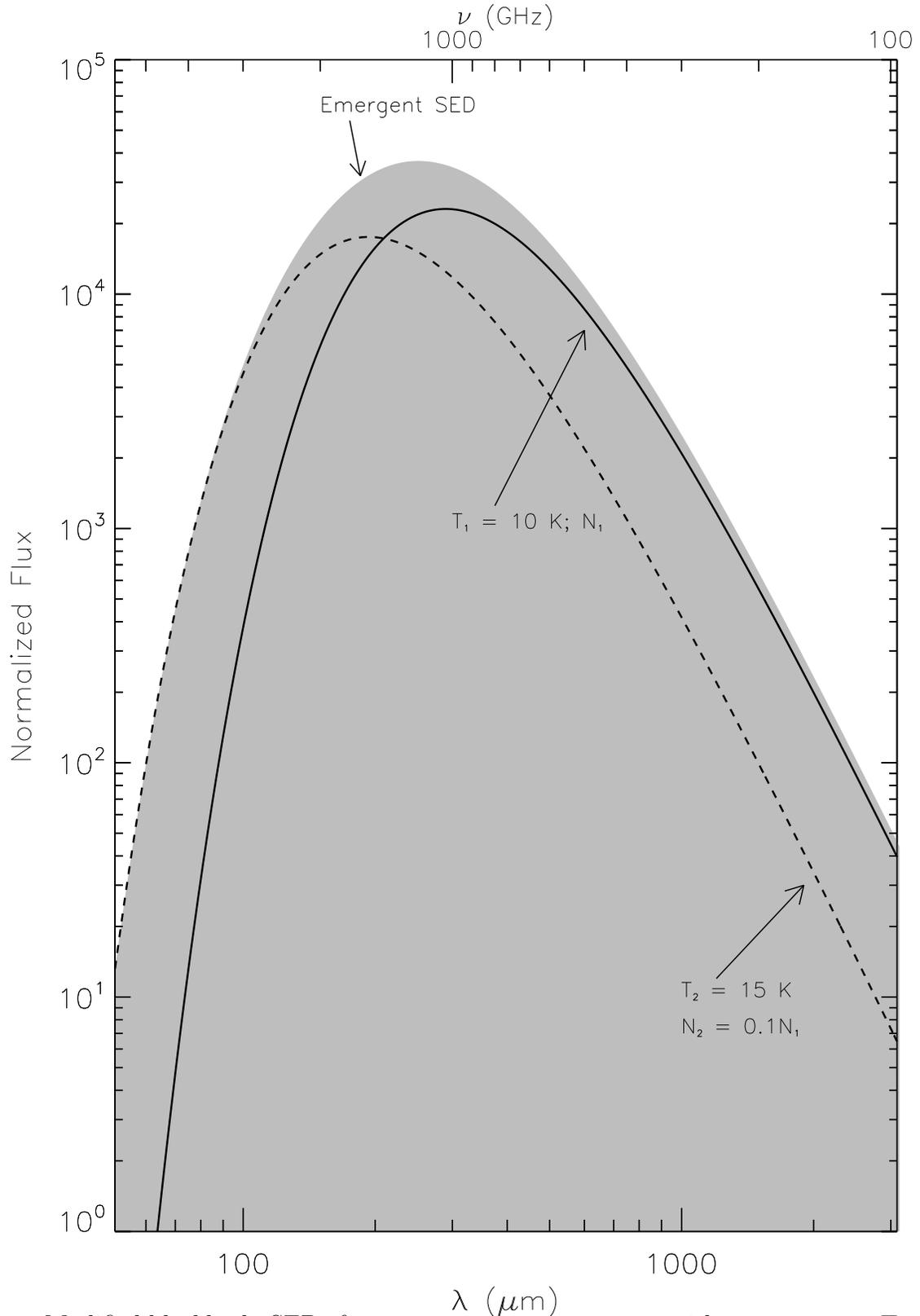}
\vspace{-0.65in}
\caption{Modified blackbody SEDs from a two-component source with
temperatures $T$= 10 K (solid) and $T$ = 15 K (dashed).  The column
density of the cooler source is a factor of 10 larger than that of the
warm source.  The boundary of the shaded region is the integrated SED,
from a line of sight containing both sources.}
\label{2compfigsed}
\end{figure}

\begin{figure}
\plotone{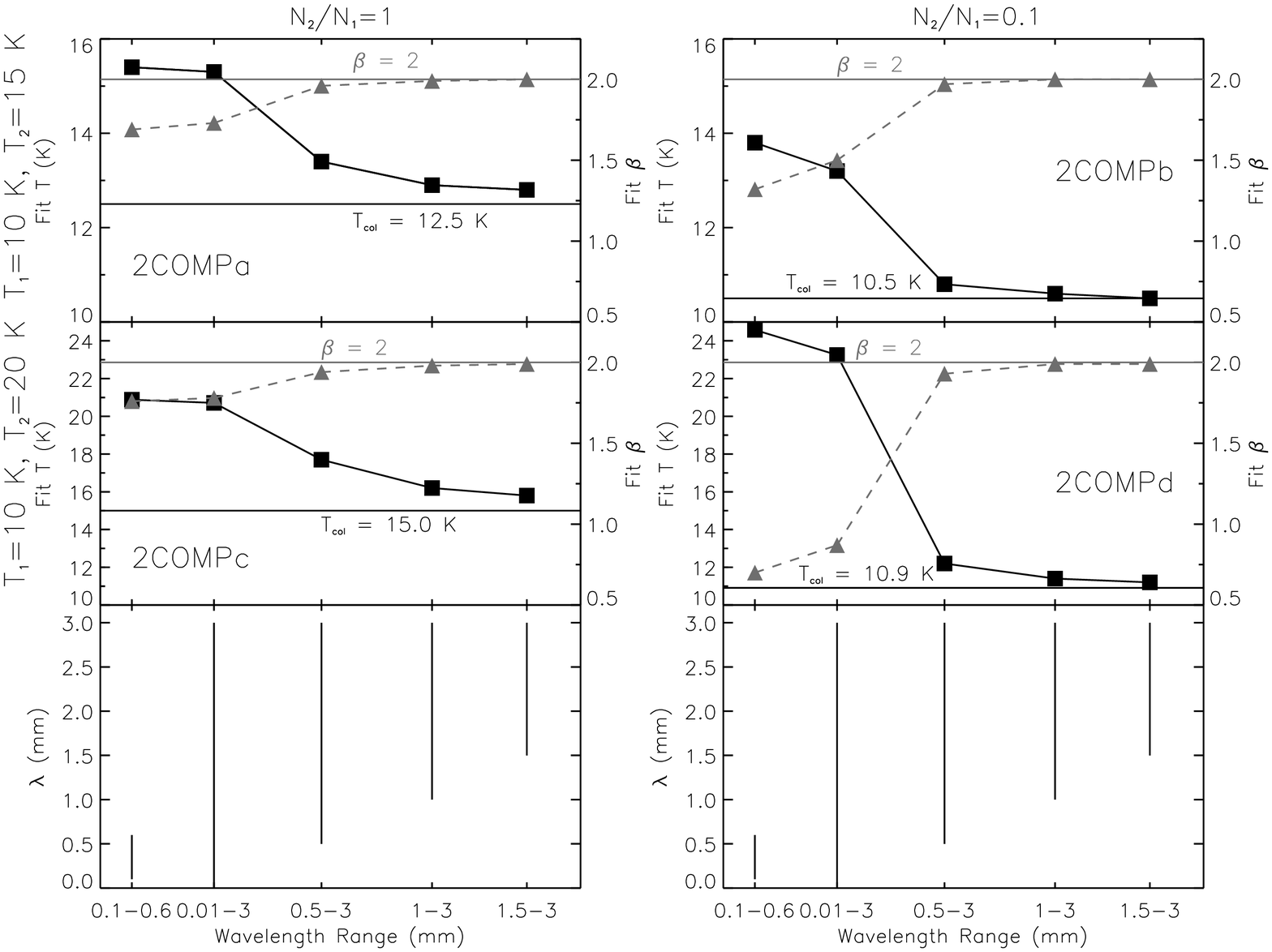}
\caption{Best fit $T$ (solid) and $\beta$ (dashed) to emergent SEDs
from two-component sources, using fluxes in different wavelength
ranges.  The top and middle rows show the best fit $T$ and $\beta$;
the left ordinate shows $T$ and the right ordinate shows $\beta$.  The
dark line shows the column temperature $T_{col}$, and the light line
indicates the $\beta$.  The bottom row shows the corresponding
wavelength ranges of the fluxes used in each fit.  The emergent SED
from ``2COMPb'' is shown in Figure \ref{2compfigsed}.}
\label{2compfig}
\end{figure}

\begin{figure}
\plotone{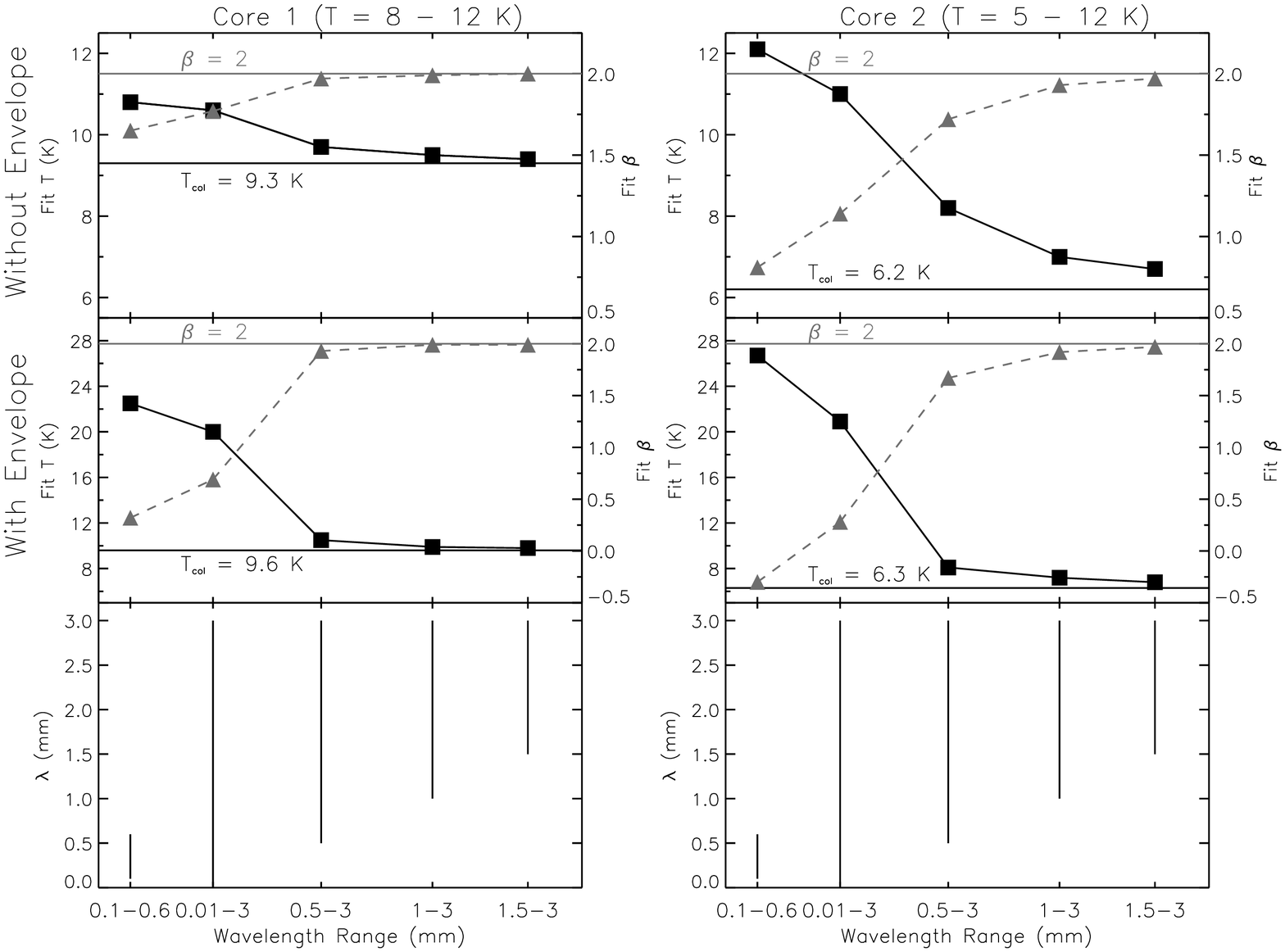}
\caption{Best fit $T$ (solid) and $\beta$ (dashed) for two cores with
and without an envelope, using fluxes in different wavelength ranges,
as in Figure \ref{2compfig}.  For Core 1, $T$ varies between 8-12 K
and $N$ between 2$\times 10^{21}$ - $1.25\times 10^{22}$ cm$^{-2}$;
For Core 2, $T$ varies between 5-12 K and $N$ between 2$\times
10^{21}$ - $1\times 10^{23}$ cm$^{-2}$.  For the Envelope, $T$ = 20 K
and $N$ = 10$^{21}$ cm$^{-2}$.  The dark line shows the column
temperature, and the light line indicates the spectral index.}
\label{modcores}
\end{figure}

\begin{figure}
\plotone{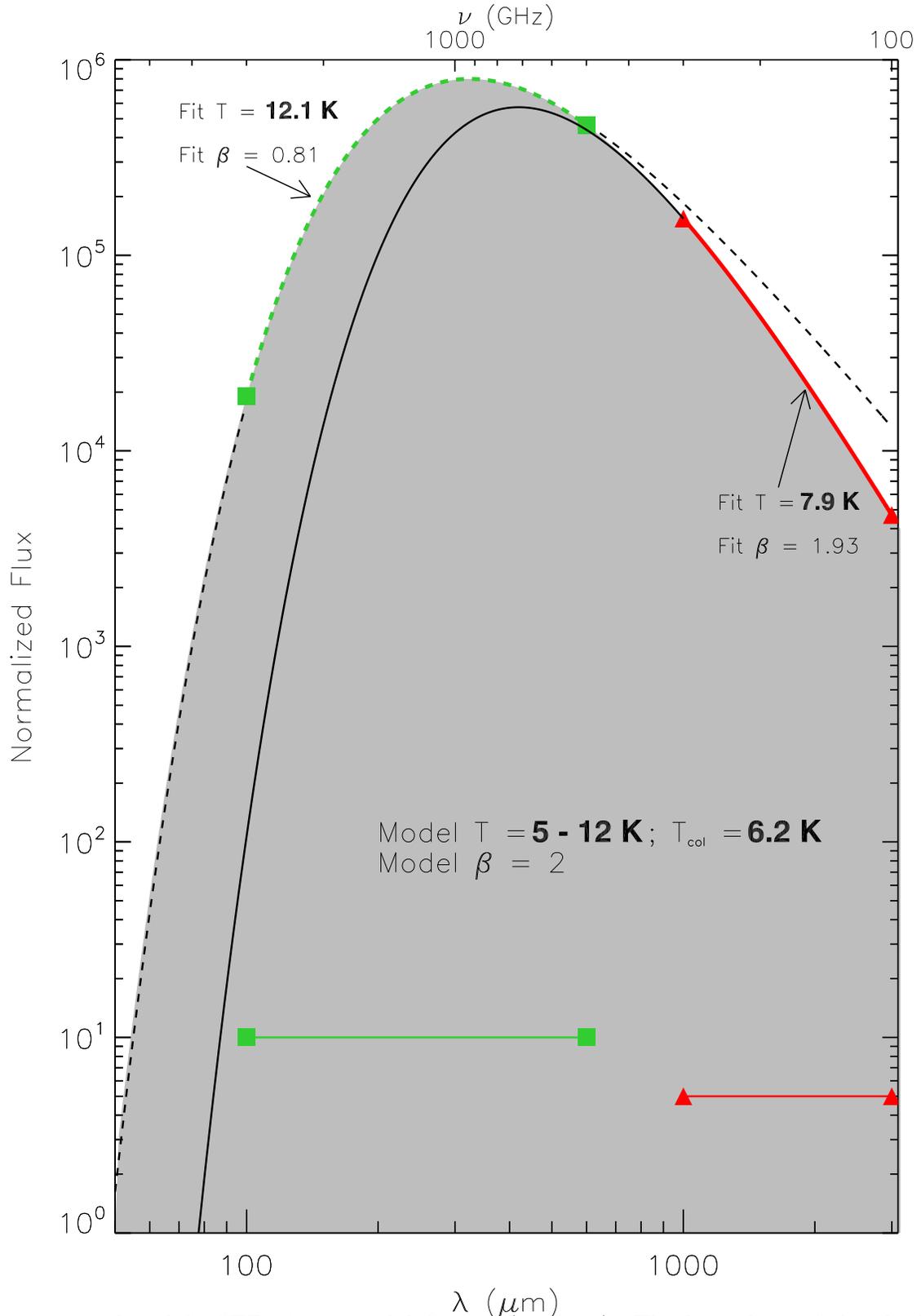}
\vspace{-0.65in}
\caption{Actual and fit SEDs from model Core 2 (see text).  The
boundary of the shaded region is the emergent SED of the core.  The
dashed SED shows the best fit to fluxes between 100 - 600 \micron\
(marked by squares).  The solid line shows the best fit to fluxes
between 1000 - 3000 \micron\ (marked by triangles).  The green and red
lines marks the extent of the wavelength ranges used in the two fits.}

\label{coresed}
\end{figure}

\begin{figure}
\plotone{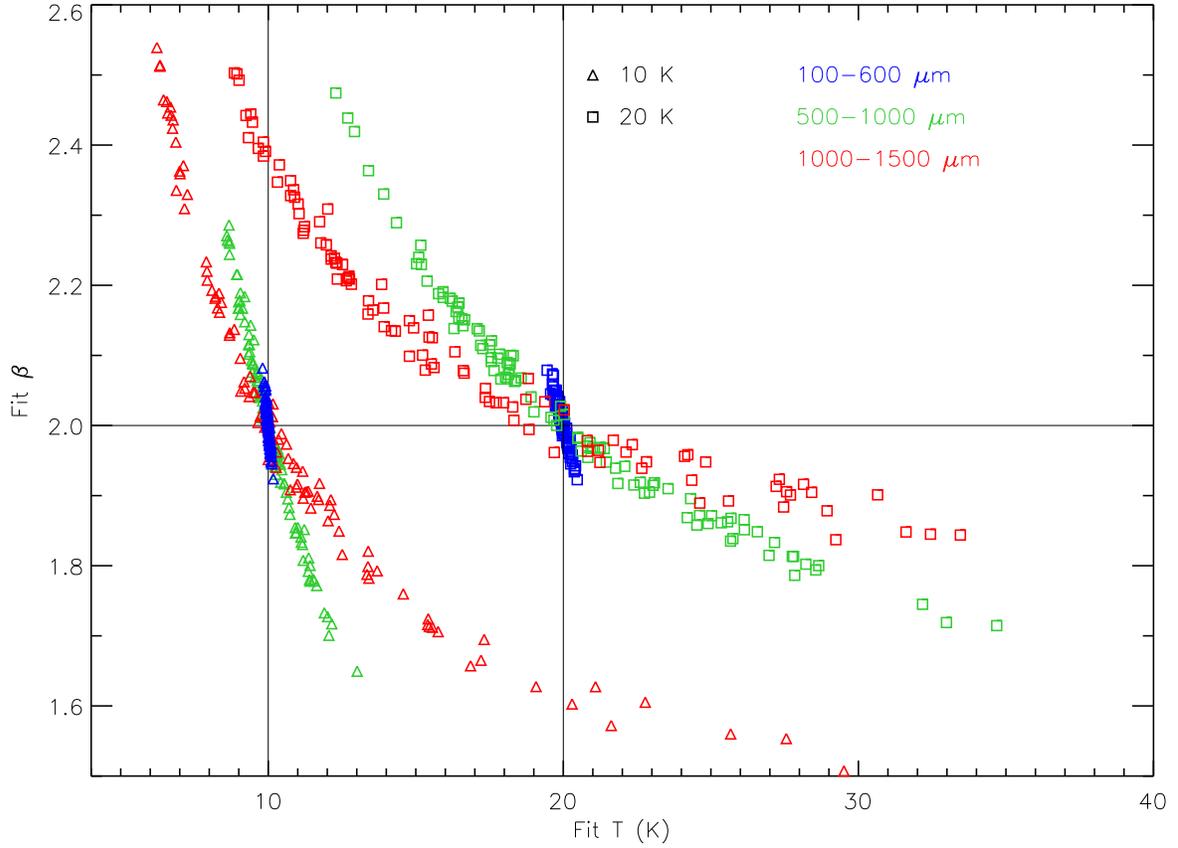}
\caption{Best fit $T$ and $\beta$ from Monte Carlo simulations of
noisy fluxes from 10 K (triangles) and 20 K (squares) isothermal
sources.  The vertical lines indicate the true source temperatures,
and the horizontal line marks the true spectral index.  Different
wavelengths fluxes were considered in each fit: 100-600 \micron\
(blue), 500-1000 \micron\ (green), and 1000-1500 \micron\ (red).
Gaussian distributed noise is added to each flux, with $\sigma$ =
5\%.}
\label{noisefig}
\end{figure}

Fits from both the 10 K and 20 K source show little scatter when only
fluxes with wavelengths 100 - 600 \micron\ are considered in the fit,
suggesting that an SED is not very sensitive to noise in the Wien
regime.  However, at longer wavelengths, there is a large spread in
the estimated $T$ and $\beta$.  Including fluxes between 500 - 1000
\micron, the fits from the 10 K source give $\beta \in 1.6$ - 2.3
(within $\sim$20\% of the source value) and $T \in$ 8 - 14 (within
$\sim$40\%).  This range increases when including fluxes between 1000
- 1500 \micron: $\beta \in 1.5 - 2.6$ (within $\sim$30\%) and $T \in 6
- 30$ (within only $\sim$200\%).  That the longer wavelength fits show
more scatter is not unexpected given the shape of an SED (see
Fig. [\ref{sedintro}]).

In the R-J tail, the SEDs from sources with different temperatures are
similar in shape, since the slope is determined primarily by $\beta$
(see Fig. [\ref{sedintro}]).  Thus, small errors in the observed
fluxes may result in inaccurate $\beta$ (and thus $T$) fits.  At
shorter wavelengths, $T$ largely determines the shape of the SED, so
small uncertainties may be insufficient to significantly alter the
temperature that best matches the observation in a fit.  For the 20 K
source in Figure \ref{noisefig}, there is less of a difference between
the spread in $\beta$ and $T$ for the fits to 500 - 1000 \micron\ and
1000 - 1500 \micron\ fluxes.  This occurs because at 20 K, the range
spanning the shorter wavelengths (500 - 1000 \micron) is already well
enough into the R-J region of the spectrum (see
Fig. [\ref{sedintro}]), so the fit is already very sensitive to noise.

The clear inverse $\beta$ - $T$ trend that emerges from SED fits to
noisy fluxes is similar to the trends found from fits to noise-free
fluxes from non-isothermal sources (as suggested by
Figs. \ref{2compfig} and \ref{modcores}).  Apparently, whenever a fit
underestimates the temperatures, the spectral index is overestimated,
and vice versa.  This effect is amplified when noisy fluxes in the R-J
regime of the spectrum are considered in the fit, and can be
understood when comparing SEDs with different values of $\beta$.
Consider a 20 K isothermal source, with $\beta$ = 2, shown in Figure
\ref{sedintro}.  Assume the source is observed at various wavelengths,
primarily at $\lambda >$ 1 mm, but that the peak is also sampled.
Further, assume the noise level in the fluxes is such that the least
squares fit preferentially obtains a $\beta$ of 1.  The peak of the
SED with $\beta$ = 1 in Figure \ref{sedintro} occurs at longer
wavelength than the peak of the $\beta$ = 2 SED, and so a fit $T$ = 20
K, with $\beta = 1$, would not reproduce the peak of the observed SED
well.  In order for the fit to reproduce the peak of the SED from the
20 K source, but with a best fit $\beta$ of 1, the best fit $T$ must
be held at a larger value than 20 K.  In general, therefore, when a
fit underestimates $\beta$, $T$ is overestimated.

We have shown that an inverse correlation between $T$ and $\beta$ can
occur due to an incorrect assumption of isothermality, or due to
intrinsic noise in the observations.  Such a trend would of course
also occur for noisy observations of non-isothermal sources.  We
discuss the combination of noise and line of sight temperature
variations when we discuss the additional limitation of including only
a small number of fluxes in $\S$\ref{obimpsec}.

\subsection{Estimating $T$ and $\beta$ with Sparse Wavelength Coverages \label{limfluxessec}}

The fitting experiments indicate that, for lines of sight with
starless-core-like temperature and density gradients, the resulting
fits produce lower temperatures and higher spectral indices as shorter
wavelength fluxes are excluded in the fit.  In all our tests so far,
we have included all wavelengths within a given range.  In practice,
however, obtaining only a (small) number of observations at different
wavelengths of a source is typically feasible.  Current (e.g.{\it
Spitzer}, {\it SCUBA}, {\it Bolocam}, {\it MAMBO}) and near future
observations (e.g. {\it Herschel} and {\it Planck}) can provide fluxes
at a number of FIR and sub-millimeter wavelengths.  However, many of
those wavebands provide fluxes at or near the peak of the integrated
SEDs of dense cores.  Thus, SED fitting may not provide accurate
estimates of the (column) temperature and spectral index.  Yet,
determining whether or not a core contains temperature variations is
instructive in itself.  Further, the temperature obtained from a fit
to long wavelength fluxes can be deemed the upper limit of the coldest
region within the core.  We thus investigate whether the
identification of temperature variations along the line of sight, as
well as an accurate estimate of the temperature limit, are still
feasible when a small number of fluxes, including those with
wavelengths near the peak of the SED, is used in the fit.  We begin by
describing the effect of fitting an SED to a limited number of
noise-free fluxes.  After that, we consider the effect of noise in
observations of sources with temperature variations, in
$\S$\ref{obimpsec}.

We construct another optically thin 2 component medium, with $T_1$ =
10 K, $T_2$ = 15 K, $N_2/N_1$ = 0.02, and $\beta$=2, similar to
``2COMPb'' in Figure \ref{2compfig}, but with a greater density
contrast.  We perform a fit assuming fluxes were obtained at 60, 100,
200, 260, 360, and 580 \micron.  The reason we choose these particular
parameters, which are realistic for innermost and outermost regions of
a core, is to assess whether any patterns emerge from fitting
different fluxes from a source with only slight temperature
variations.  We choose fluxes at wavelengths near the peak for more
direct comparisons with real observations of cores.  The column of
panels in Figure \ref{corefigs}(a) shows the fitting results.  The
bottom panel of Figure \ref{corefigs}(a) indicates the observed
wavelengths, along with the emergent SED from the 2 component source.
There are no significant differences in the fitting results when
excluding the 60 \micron\ flux.  However, when the additional flux at
100 \micron\ is excluded, the best fit temperature decreases and
spectral index increases.  We were unable to obtain a good fit with
only three fluxes; we have found that a minimum of four wavelengths is
necessary to obtain a fit with three free parameters.\footnote{As we
discuss in $\S$\ref{comparesec}, a good fit can be found with only
three data points if one (or more) of the fit parameters ($\beta$,
$T$, or $N\kappa_0$) is held fixed.}

Excluding short wavelength fluxes, even with a limited number of
observations in the Wien regime, results in a lower estimate of $T$
and a higher estimate of $\beta$.  We found the same trend regardless
of what wavelengths were considered in the fit, and for a variety of
systems with an inverse relationship between the column density and
temperature.

\subsection{Implications to Recent Observations \label{obimpsec}}

\begin{figure}
\plotone{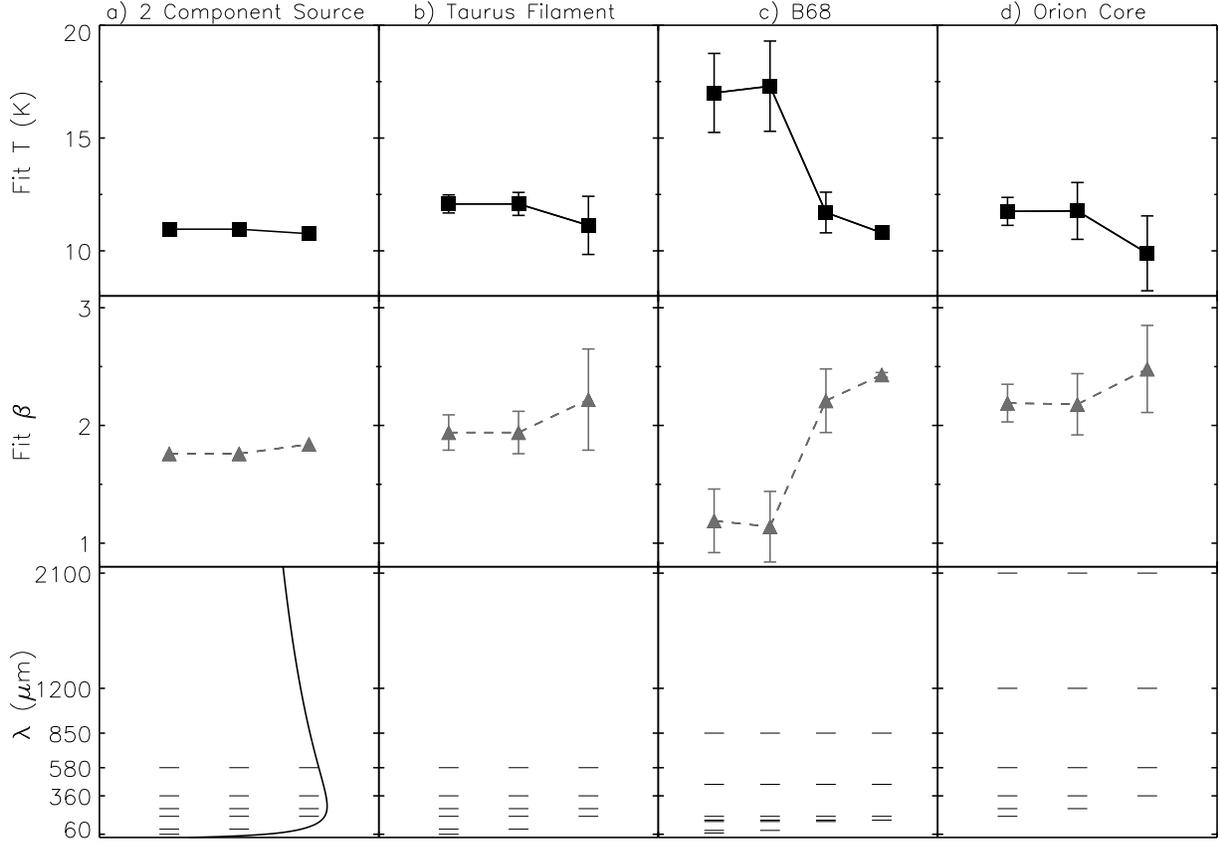}
\caption{Best fit $T$ (top row) and $\beta$ (middle row) to small
number of observed fluxes, marked by the line segments in the bottom
row.  a) A 2 component source (with $T_1$ = 10 K, $T_2$ = 15 K, and
$N_2/N_1$=0.02), observed at $\lambda$ = 60, 100, 200, 260, 360, and
580 \micron.  The bottom panel also shows the emergent SED from this
source, with the abscissa corresponding to log($S_\nu$); b) Filament
in Taurus, observed at $\lambda$ = 60, 100, 200, 260, 360, and 580
\micron\ by \citet{Stepniketal03}; c) B68, observed at $\lambda$ = 70,
90, 160, 170, 200, 450, and 850 \micron\ by \citet{Kirketal07}; and d)
Core in Orion, observed at $\lambda$ = 200, 260, 360, 580, 1200, and
2100 \micron\ by \citet{Dupacetal01}.}
\label{corefigs}
\end{figure}

We now perform a similar tests to published observations of cold star
forming regions, beginning with the observations presented by
\citet{Stepniketal03}.  They analyzed a filament in the Taurus
molecular cloud using 60, 100, 200, 260, 360, and 580 \micron\ fluxes
observed from {\it IRAS} and the balloon borne experiment PRONAOS/SPM.
After subtracting off emission attributed to the surrounding envelope,
they fit the 6 data points to obtain an estimate of the temperature
and spectral index of the filament.  We carry out the exact same
procedure, but also perform fits excluding the short wavelength
fluxes.  The fitting results are shown in Figure \ref{corefigs}(b).
As with the two-component model shown in Figure \ref{corefigs}(a), the
exclusion of the 60 \micron\ flux does not alter the fit.  However,
additionally excluding the 100 \micron\ flux decreases the fit $T$ by
$\sim$ 1 K, and increases $\beta$ from 1.98 to 2.13.  This variation
is rather similar to the simple 2 component fit, indicating that there
might be a temperature variation within the filament, and that the
actual value of the spectral index is greater than 2.1 (if the
spectral index itself is constant in the filament).  A value of 11.13
$\pm$ 1.29 K can be assigned as an upper limit for the column
temperature of the filament.  The interpretation of a temperature
variation within the filament cannot be definitive, however, due to
the large uncertainties.  Within the uncertainties, the fit $T$ and
$\beta$ are constant regardless of which fluxes are considered in the
fit.  More observations, preferably at longer wavelengths, are
necessary to confidently determine whether there are temperature
variations within the filament itself, and for an accurate estimate of
the spectral index.

\citet{Kirketal07} also employed SED fitting to {\it Spitzer} and {\it
ISO} observations to estimate the temperatures of numerous cores.  We
analyze their fluxes in a similar fashion.  Though a few of the cores
did not show definitive temperature drop as shorter wavelength fluxes
were excluded, the well studied core B68 shows a clear drop in
temperature and an increase in the spectral index after short
wavelength fluxes are excluded.  The fitting results for B68 is shown
in Figure \ref{corefigs}(c).\footnote{Using all 7 of their fluxes,
\citet{Kirketal07} obtain a best fit $T$=12.5 K; the difference
between their value and ours arises because they kept $\beta$ at a
fixed value of 2.  Our results agree when $\beta$ is fixed at that
value in our fits.}  The best fit temperatures decreases by $\sim$ 5 K
when the 70 and 90 \micron\ fluxes are excluded in the fit.  There is
an additional 1 K temperature drop when the 160 \micron\ flux is
excluded.  There is also a corresponding increase in $\beta$ from
$\sim$ 1.2 to $\sim$2.4.  One interpretation of such a high spectral
index is that the dust grains are covered by icy mantles
\citep[e.g.][]{Kuanetal96}; this interpretation would be reasonable
for B68, where there is significant molecular depletion onto dust
grains \citep{Berginetal02}.  The variations in the spectral index and
temperature estimates when short wavelength fluxes are excluded in the
fit, along with the relatively small error bars, strongly suggests
that there are dust temperature (and corresponding inverse density)
variations within B68.  We thus assign an upper limit of 10.8 $\pm$
0.1 K for the coldest region within B68.  Our fits also suggest that
the spectral index of dust in B68 \apgt\ 2.4, if that property is
constant throughout the core.

We perform the same test of the data presented by \citet{Dupacetal01}.
They estimated the temperature and spectral indices of regions in the
Orion complex.  One of the regions appears to be a dense core without
a central source, referred to as ``Cloud 2'' by \citet{Dupacetal01}.
Figure \ref{corefigs}(d) shows our fitting results.  For this core,
longer wavelength data at 1.2 and 2.1 mm are available.  In this case,
the exclusion of short wavelength data reduces the temperature by
$\sim$ 2 K.  The spectral index also increases, from $\sim$2.2 to
$\sim$2.5.  These results are also suggestive of temperature
variations within the core, but due to the relatively large
uncertainties, an isothermal description cannot be ruled out.

We have found that the systematic exclusion of short wavelength data
for dense cores results in lower best fit temperatures, and higher
best fit spectral indices.  A trend of decreasing $\beta$ with
increasing $T$ has been put forward as a physical property of dust
grains in the ISM by \citet{Dupacetal03}.  They find such a trend by
fitting equation (\ref{fd}) to observed fluxes primarily from
PRONAOS/SPM, corresponding to wavelengths (in the range 100 - 600
\micron) near the peak of the SEDs emitted by dust.
\citet{Dupacetal03} argue that line of sight temperature variations
only result in a slight variation of $\beta$ with $T$, and that
unrealistically high density contrasts ($\sim$ 100 $\times$) are
necessary to reproduce the magnitude of the inverse $T$ - $\beta$
correlation \citep[see also][]{Dupacetal02}.  We find that temperature
variations in realistic cores with a uniform spectral index would show
an inverse $T$ - $\beta$ relationship when observed at wavelengths
near the peak of the emergent SED.  Besides resulting in erroneous
$\beta$ estimates, the best fit parameters vary as short wavelength
fluxes are excluded in the fit, with the best fit $\beta$ approaching
the correct value when only long wavelengths are considered.  Further,
we also find that modest errors, as low as 5\%, in observed fluxes
from isothermal sources lead to erroneous $\beta$ and $T$ estimates.
The trend in the points in a $T$ - $\beta$ plane from a number of fits
to noisy fluxes also show such an inverse correlation
(Fig. [\ref{noisefig}]).  It would thus be informative to compare the
form of the $T$ - $\beta$ relationship we find with that of
\citet{Dupacetal03}.  We note that though we only consider objects
with $T\le$ 20 K in this work, we investigate the derived inverse $T$
- $\beta$ relationship due to noise in observations of isothermal
sources with temperatures up to 100 K in an accompanying paper
\citep{Shettyetal09}.

\subsubsection{A True Inverse $T$ - $\beta$ Correlation?}

To further investigate the derived inverse $T$ - $\beta$ correlations,
we again consider the two-component medium ``2COMPd,'' with
temperatures $T_1$ = 10 K, $T_2$ = 20 K, and a column density ratio
$N_2/N_1$ = 0.1.  As shown in Figure \ref{2compfig}, a modified
blackbody SED fit results in a $T$ estimate of 23.3 K and a $\beta$
estimate of 0.23, using fluxes measured between $\lambda \in$ 10 -
3000 \micron.  To compare with realistic observations, we perform the
fit to data sets each containing fluxes at 5 wavelengths.  Two sets of
wavelengths are considered: one with fluxes near the peak of the SED
at 100, 200, 260, 360, and 580 \micron, and the other with fluxes in
the R-J tail of the SED at 850, 1100, 1200, 1500, and 2100 \micron.
In order to account for the effect of a $\sim$5\% uncertainty in the
observations, due to noise or calibration errors, for example, each
flux is multiplied by a random value drawn from a Gaussian
distribution with a mean of 1.0 and dispersion of 0.05.  We generate
100 sets of fluxes in this manner, and fit equation \ref{fd} to each
set of fluxes, as in the numerical experiments of isothermal sources
presented in $\S$\ref{isonoise}.

\begin{figure}
\plotone{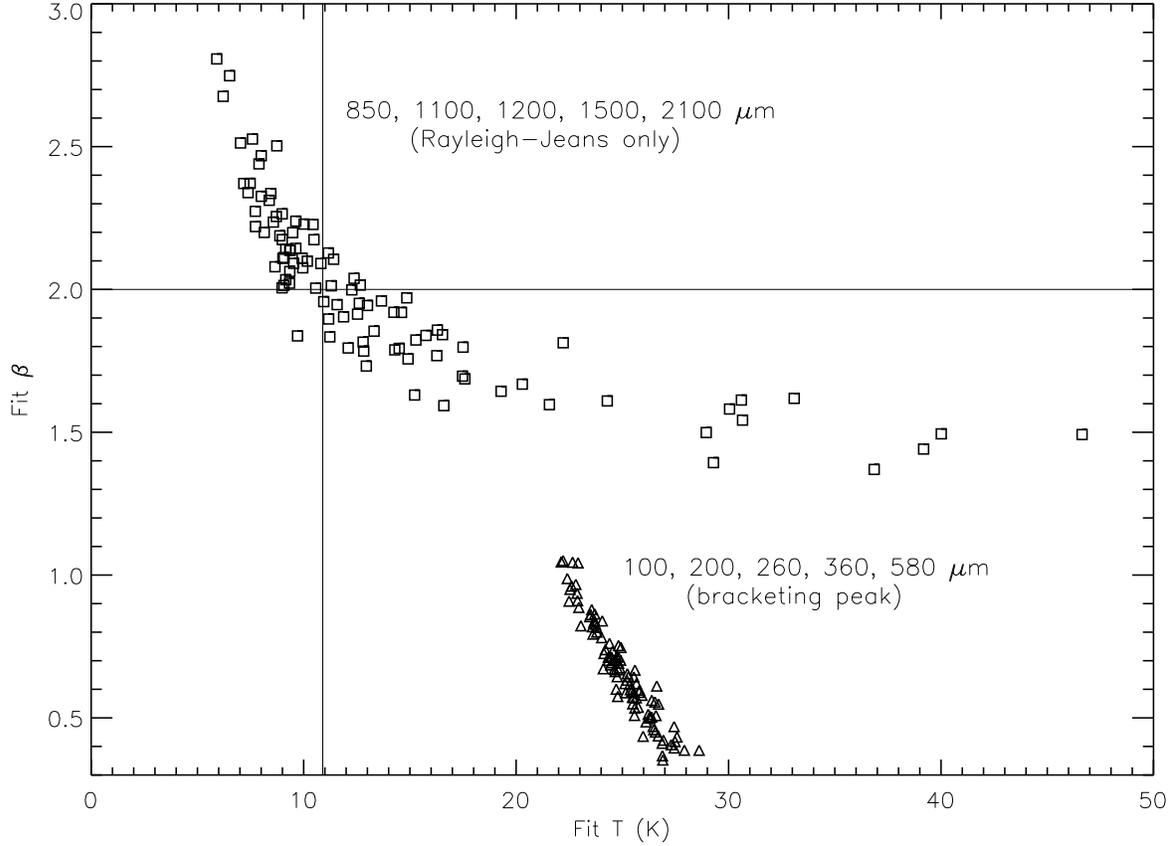}
\caption{Best fit $\beta$ and $T$ to observed fluxes from the 2
component source ``2COMPd'' (with $T_1$ = 10 K, $T_2$ = 20 K, and
$N_2/N_1$=0.1; see Fig. [\ref{2compfig}]).  Five fluxes are used in
each fit: 850, 1100, 1200, 1500, and 2100 \micron\ (squares), or 100,
200, 260, 360, and 580 \micron\ (triangles).  Each flux includes a
small (Gaussian distributed) random component, with $\sigma=$ 5\%.
The lines indicate the model parameters $\beta$ = 2 and column
temperature = 10.9 K.}
\label{betaT}
\end{figure}

Figure \ref{betaT} shows the fitting results from these simulated
observations.  The fits to the observed fluxes at long wavelengths
show a scatter in the $T$ - $\beta$ relation, due to the uncertainties
in the fluxes.  The fits at shorter wavelengths do not show as much
scatter, but the best fit $\beta$ and $T$ are very poor estimates of
the true $\beta$ or the column temperature of 10.9 K.  As previously
discussed, this inaccuracy at short wavelengths results because the
emergent SED from a non-isothermal source is not well fit by a single
power law modified blackbody spectrum.

\citet{Dupacetal03} observe a variety of sources, including dense
cores as well as much warmer sources.  After fitting modified
blackbody SEDs to the observed fluxes, with wavelengths $<$ 600
\micron, they find that a hyperbolic form in $\beta(T)$ represents the
fits well.  They investigated the effect of noise in a model with a
range of uncorrelated $T$ - $\beta$ pairs, and concluded that such a
model is inconsistent with the observed data.  The shape of the $T$ -
$\beta$ correlation in Figure \ref{betaT} from a model with a single
$\beta$ is remarkably similar to that shown in \citet{Dupacetal03}.
This suggests that an inverse, and possibly even hyperbolic-shaped,
$T$ - $\beta$ relationship is not necessarily due to real variations
in the dust spectral index with dust temperature.  The relationship
may simply be due to temperature variations along the line of sight,
along with uncertainties in the observed fluxes.  At the short
wavelengths considered by \citet{Dupacetal03}, for the warmer sources
the observed wavelengths may indeed fall in the R-J part of the
spectrum.  For these sources, the ``short wavelength'' fits will be
analogous to the ``long wavelength'' set in Figure \ref{betaT}.  For
sources that are dense cores, if they are not isothermal, the SEDs at
those short wavelengths are not well fit by equation (\ref{fd}); a fit
would produce a large $T$ estimate, relative to the column
temperature, and would underestimate $\beta$.

For isothermal sources ($\S$\ref{isonoise}), the addition of noise to
the observed fluxes further degrades the parameter estimates.  Our
analysis indicates that the SED fits to fluxes with $\lambda <$ 600
\micron\ from various warm isothermal sources with $T$ \apgt 60 K may
all produce similar $T$ and $\beta$ estimates \citep{Shettyetal09}.
Uncertainties in the observed fluxes of starless cores may be
responsible for some of the scatter in the $T$ - $\beta$ diagram shown
by \citet{Dupacetal03}.  However, all of the spread is likely not a
consequence solely of noise, since \citet{Dupacetal03} observe a
variety of sources.  One possibility is that the spectral index varies
within a source, which is a situation we do not model (see Table
\ref{overv}).  The emergent SED from such sources will of course be
more complicated, for which alternative analysis techniques may
provide better parameter estimates.  We note that the points in Figure
\ref{betaT} would simply be systematically offset had we considered a
source with a different (constant) spectral index from $\beta$ = 2
\citep{Shettyetal09}.  Thus, had we included multiple sources with
different (constant) values of $\beta$, the $T$ - $\beta$ diagram
would be further populated.  Though we cannot exclude the possibility
that the spectral index of dust decreases with increasing temperature,
we have shown that simple power-law-modified-blackbody fits to
observed data can result in misleading $T$ - $\beta$ relationships
which appear like those sometimes claimed to be of physical origin.

We have demonstrated that the assumption of isothermality can lead to
significant errors in estimates of $T$ and $\beta$ from SED fits.
Noise contributes to additional uncertainty in the estimated
parameters.  In the next section, we describe the effect of noise on a
different method commonly used to estimate $T$ and $\beta$, by means
of flux ratios, focusing on isothermal sources.  We then compare the
flux ratio estimates to those derived from SED fitting for sources
with line of sight temperature variations.

\section{Flux Ratios \label{flxsec}}

An alternative to full SED fitting for estimating the temperature is
through the use of ratios of observed fluxes.  For a given
observation, the known quantities in equation (\ref{fd}) are the flux
density $S_\nu$ and the beam size $\Omega$.  With two observations
(smoothed to a uniform resolution) at different frequencies $\nu_1
{\rm \,and\,} \nu_2$, corresponding to wavelengths $\lambda_1 {\rm
\,and\,} \lambda_2$, the ratio of $S_{\nu_1}/S_{ \nu_2}$ produces an
equation where the two unknown quantities are $T$ and $\beta$:
\begin{equation}
\frac{S_{\nu_1}}{S_{\nu_2}} =
\left(\frac{\lambda_2}{\lambda_1}\right)^{3+\beta} \frac{\exp(\lambda_T/\lambda_2)-1}{\exp(\lambda_T/\lambda_1)-1},
\label{2obsratio}
\end{equation}
where $\lambda_T = hc/kT$.  The main assumptions made to derive this
equation are that $T$ and $\beta$ are constant along the line of
sight.  If $\beta$ is known {\it a priori} (or otherwise assumed to be
known), then equation (\ref{2obsratio}) can be used to estimate the
temperature from only two observations
\citep[e.g.][]{Krameretal03,SchneeGoodman05,WardThompsonetal02,Schlegeletal98}.

Including a third observation at frequency $\nu_3$, corresponding to
wavelength $\lambda_3$, taking ratios using all three flux densities
produces
\begin{subequations}\label{ratio1_2}
\begin{align}
\log \left( \frac{S_{\nu_1}}{S_{\nu_2}} \right)\log\left(\frac{\lambda_3}{\lambda_2}\right) - \log \left( \frac{S_{\nu_2}}{S_{\nu_3}} \right)\log\left(\frac{\lambda_2}{\lambda_1}\right) = \label{ratio1} \\
\log \left[ \frac{\exp(\lambda_T/\lambda_2)-1}{\exp(\lambda_T/\lambda_1)-1} \right]\log\left(\frac{\lambda_3}{\lambda_2}\right) - \log \left[ \frac{\exp(\lambda_T/\lambda_3)-1}{\exp(\lambda_T/\lambda_2)-1} \right]\log\left(\frac{\lambda_2}{\lambda_1}\right) \label{ratio2}.
\end{align}
\end{subequations}
The advantage of using this equation to estimate the temperature is
that no assumption for the value of $\beta$ is required, though
$\beta$ is assumed to be constant along the line of sight.  A similar
equation can be derived for observations at four wavelengths.
However, as we discuss in $\S$\ref{discsec}, in that case a direct fit
of equation (\ref{fd}) is reliable.  We will hereafter refer to the
left hand side of equation (\ref{ratio1_2}) as the ``flux ratio,'' and
the right hand side as the ``analytic prediction.''

\citet{Schneeetal07} used fluxes from observations of the starless
core TMC-1C at 450, 850, and 1200 $\mu$m in equation (\ref{ratio1_2}).
They found that the errors in the observations would have to be \aplt
2\% in order to accurately estimate the temperature (of an isothermal
source).  The goal of that study, besides mapping the temperature, was
also to map the spectral index and column density.  Once an estimate
for the temperature is obtained, the spectral index can be estimated
through the use of the ratio of any two fluxes as:
\begin{equation}
\beta = \log \left[ \frac{S_{\nu_1}}{S_{\nu_2}}\frac{\exp(\lambda_T/\lambda_1)-1}{\exp(\lambda_T/\lambda_2)-1} \right] / \log \left( \frac{\lambda_2}{\lambda_1} \right) - 3
\label{betaeqn}
\end{equation}

In principle one can also estimate the column density $N$, modulo
$\kappa_0$, once $T$ and $\beta$ are derived from equations
(\ref{ratio1_2})-(\ref{betaeqn}) \citep[e.g.][]{Schneeetal06}.  As
discussed in $\S$\ref{fitsec}, however, additional assumptions for the
opacity and/or the dust-to-gas ratio may be required if extinction
observations are unavailable.  Since our focus is on the temperature
and spectral index, we do not consider those assumptions here.

\subsection{Determining Temperatures from Two Fluxes \label{2wave}}

We begin our analysis of the flux ratio method by considering equation
(\ref{2obsratio}) to estimate the temperature, given two fluxes at
different wavelengths.  We consider both an isothermal source and a
two-component source (as in $\S$\ref{2comsec}).  We then compare the
fluxes at two chosen wavelengths to the analytical prediction from the
right-hand-side of equation (\ref{2obsratio}).  We analyze the effect
of noise, as well as of an incorrect assumption of $\beta$, on the
temperature estimate.  For the analysis including noise, we add a
random component drawn from a Gaussian distribution with a chosen
dispersion to the flux, and then use these ``noisy'' fluxes in
equation (\ref{2obsratio}).  We repeat this simple experiment 10,000
times to obtain better statistics on the temperature estimates.

Table \ref{2fluxes} shows the temperature estimates using observations
with different noise levels, and assuming different values of $\beta$.
Column (1) shows the wavelengths of the two fluxes.  Column (2) shows
the assumed value of $\beta$ in equation (\ref{2obsratio}); the
spectral index of the model source is 2.0.  Column (3) indicates the
level of noise added to the fluxes.  Column (4) shows the true column
temperature; for the isothermal source, the column temperature is just
the actual source temperature.  The last column gives the derived
temperature.  For the fluxes that are altered by noise, we show the
1$\sigma$ distribution in the estimated temperatures.

For an isothermal source, the shorter wavelength pair is less
sensitive to noise and/or errors in the assumed value of $\beta$.  For
the two-component source ``2COMPd'' (see Fig. [\ref{2compfig}])
without noise, the longer wavelength pair gives a (slightly) more
accurate measure of the column temperature than the shorter wavelength
pair, when the correct value of $\beta$ is assumed.  However, when
modest levels of noise are included in the fluxes, the long wavelength
pair produces temperature estimates that deviates from the true
temperature (or column temperature) more than the short wavelength
pair.  Additionally, the longer wavelength pair is more sensitive to
the assumed value of $\beta$.

These trends are as expected given the shape of the modified blackbody
(Fig. \ref{sedintro}).  As we demonstrated in $\S$\ref{fitsec} using a
direct SED fit, fluxes in the R-J part of the spectrum provide more
accurate $T$ and $\beta$ estimates, though the fits are more sensitive
to noise.  Similarly, the flux ratio involving wavelengths in the R-J
part of the spectrum is more sensitive to the assumed value of the
spectral index, as well as noise, indicating that short wavelength
fluxes are preferable.  However, when large temperature gradients are
present (see $\S$\ref{2comsec} and $\S$\ref{corevarT}), the
temperature estimate from the short wavelength pair will deviate
significantly from the column temperature.

For isothermal sources, the source temperature, as well as the assumed
value of $\beta$ determines how well fluxes at different wavelengths
could recover the temperature.  For noise free observations of a 10 K
source at 450 and 850 \micron, and when $\beta$ is assumed to be
within 5\% of the source value, $T$ can be recovered within 10\%.  The
accuracy of the $T$ estimate degrades to $\sim$20\% with 1200 and 2100
\micron\ observations.  For warmer sources ($T>> 10$ K), 450 and 850
\micron\ fluxes would be well into the R-J part of the spectrum.
These fluxes would thus be more sensitive to noise and $\beta$ than
those 450 and 850 \micron\ fluxes from a 10 K source.  Lower
temperature sources always give better (i.e. less uncertain)
temperature estimates.  The flux ratio method, however, still requires
a reasonable assumption for $\beta$.

For non-isothermal sources, the estimates become more uncertain.  As
shown in Table \ref{2fluxes}, low noise levels and an accurate
assumption of $\beta$ can reasonably recover the column temperature of
a simple 2 component medium.  Since equation (\ref{2obsratio}) is
derived from the isothermal assumption, the estimated temperature
becomes less accurate for more complex sources, even when comparing
with the column temperature.  In our subsequent analysis of flux
ratios, we will hereafter concentrate on deriving temperatures of
isothermal sources.

\subsection{Determining Temperatures from Three Fluxes \label{noisesec}}

\begin{figure}
\plotone{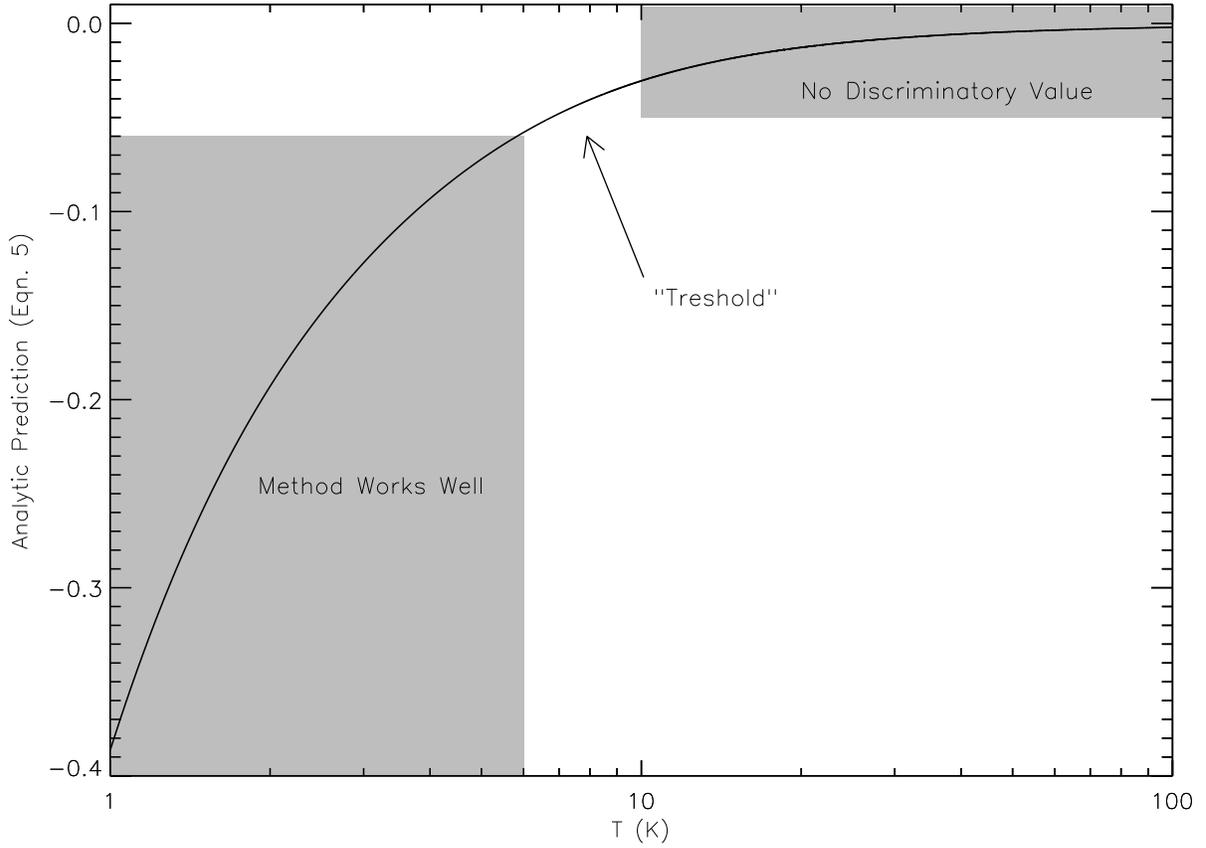}
\caption{Analytic prediction $\log \left[
\frac{\exp(\lambda_T/\lambda_2)-1}{\exp(\lambda_T/\lambda_1)-1}
\right]\log\left(\frac{\lambda_3}{\lambda_2}\right) - \log \left[
\frac{\exp(\lambda_T/\lambda_3)-1}{\exp(\lambda_T/\lambda_2)-1}
\right]\log\left(\frac{\lambda_2}{\lambda_1}\right)$ (RHS in
eqn. [\ref{ratio1_2}]) for three observations at $\lambda_1=$450,
$\lambda_2=$850, and $\lambda_3=$1200 \micron.}
\label{lookupval}
\end{figure}

We next investigate the accuracy in estimating temperatures from
observations at 3 wavelengths, using Equation (\ref{ratio1_2}).  We
first consider observations at 450, 850, and 1200 $\mu$m, the three
``popular'' wavelengths used in the \citet{Schneeetal07} study.
Figure (\ref{lookupval}) shows the analytic prediction used to
determine T from expression (\ref{ratio2}) for temperatures between 1
- 100 K.  At temperatures $T$ \aplt\ 7 - 10 K, the analytic prediction
is very sensitive to the temperature.  Thus, even though errors in the
fluxes will produce an inaccurate value in expression (\ref{ratio1})
for comparison with the analytic prediction of expression
(\ref{ratio2}), the derived temperature will still be close to the
actual temperature of the emitting medium.  At temperatures $T$ \apgt\
7 - 10 K, however, the analytic prediction is not very sensitive to
the temperature.  Thus, even small errors in the flux, due to noise
and/or other observational uncertainties, will result in grossly
erroneous temperature estimates.

To investigate the effect of noise on the determination of temperature
using equation \ref{ratio1_2}, we have run a number of Monte-Carlo
simulations.  In these simulations, the emergent flux of a source at
constant temperature (eqn. [\ref{fd}]) is modified by some chosen
level of Gaussian noise, representing random errors in real
observations.  These ``observed'' fluxes at three wavelengths are used
in expression (\ref{ratio1}) to compare with the analytic prediction
of expression (\ref{ratio2}).

\begin{figure}
\plotone{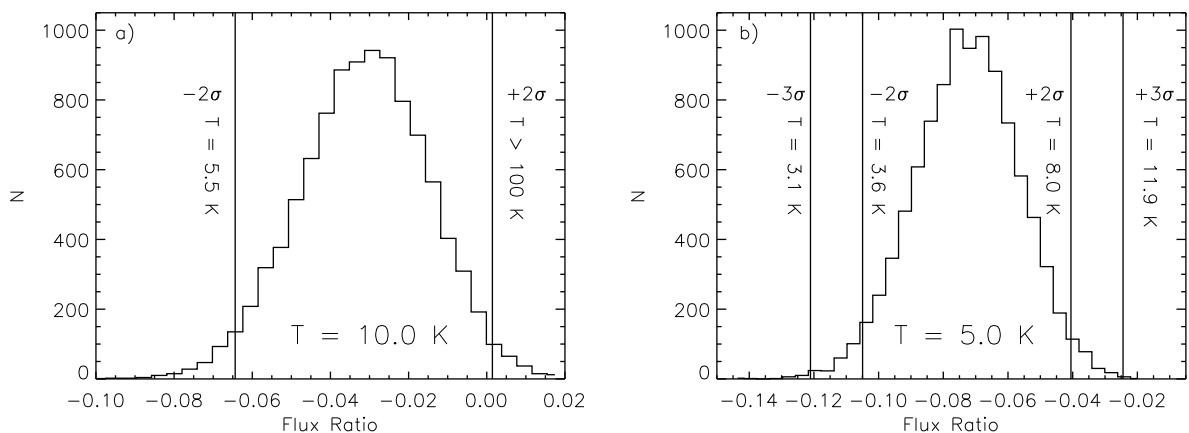}
\caption{Histogram of flux ratios (LHS of eqn. [\ref{ratio1_2}]), including
noise, for sources at (a) 10 K and (b) 5 K.  Noise levels were set at
12\%, 4\%, and 10\% for the 450, 850, and 1200 $\mu$m observations,
respectively.  Mean flux ratio corresponds to (left) 9.8 K, and
(right) 5.0 K.  Lines on the 10 K histogram show the $\pm 2\sigma$
level, corresponding to temperatures of $>$ 100 K and 5.5 K (see
Fig. \ref{lookupval}).  Lines on the 5 K histogram show $\pm 2\sigma$
levels, corresponding to 8.0 K and 3.6 K, and $\pm 3\sigma$,
corresponding to 11.9 K and 3.1 K.}
\label{hist1}
\end{figure}

The uncertainties in the observations of \citet{Schneeetal07} were
estimated at 12\%, 4\%, and 10\% for the 450, 850, and 1200 $\mu$m
observations, respectively.  We use those uncertainties in our first
simulated observations.  We ``observe'' the extended source at 10,000
positions (or, equivalently, 10,000 times at a single location on the
sky) at those three wavelengths.  The flux ratios, using the noisy
fluxes in expression (\ref{ratio1}), from two sources at 10 K and 5 K
are shown in Figure \ref{hist1}.  The mean values of the flux ratio
recovers the true temperature reasonably accurately.  Also marked are
the $\pm 2\sigma$ levels (as well as the $\pm 3\sigma$ levels for the
5 K source).  At 10 K, even at the $+2\sigma$ level, the flux ratio
does not lie in the 1 - 100 K range of the analytic prediction shown
in Figure \ref{lookupval}.  However, for the source at 5 K, at the
$3\sigma$ level the derived temperature only differs from the true
temperature by a factor of $\sim2.4$.  At 10 K (and higher)
temperatures, Figure \ref{lookupval} shows that the analytic
prediction does not vary much with temperature, so any error in the
flux ratio will correspond to a temperature that deviates
significantly from the true temperature.  For a 5 K source, the
analytic prediction varies significantly with slight variations in
temperature, so errors in the flux ratio will still produce reasonably
accurate temperature estimates.  This test has shown that with
observations at 450, 850, and 1200 $\mu$m, one can only be confident
in the ratio method if the estimated temperatures are \aplt 5 K.  For
temperatures greater than the turn-over temperature in Figure
\ref{lookupval} ($\sim$ 7 - 10 K), the derived temperatures cannot be
deemed accurate.

The temperature at which analytic prediction shifts from a strong
temperature dependence to a weak temperature dependence is determined
by the particular wavelengths used in expression (\ref{ratio2}).  In
this sense, the ideal set of three wavelengths would shift the
turn-over temperature to much higher values.  In order to test the
sensitivity of the turn-over temperature to the particular values of
wavelengths, we have run a series of Monte-Carlo simulations as
described above; we varied both the set of 3 wavelengths, as well as
the (constant) temperature of the source.

In order to locate the turn-over temperature, we define a threshold
temperature $T_{th}$ such that a $\pm3\sigma$ range in the flux ratios
corresponds to estimated temperatures $T_{est}$ that are within a
factor of 2 of the actual source temperature $T_0$.  For example,
consider a source at $T_0$ = 21 K observed at three given wavelengths.
If the range of derived temperatures included in the $\pm3\sigma$
level of the flux ratio includes temperatures $>$ 42 K, then we know
that $T_{th} < 21$ K for that set of wavelengths.  If we then run
another simulation on a source with $T_0$ = 20 K, with the same three
wavelengths, and find that the maximum derived temperature in the
3$\sigma$ range is $T_{est} < 40$ K, then we set $T_{th} = 20$ K for
this set of three wavelengths.  Our definition of $T_{th}$ is
arbitrary, and can of course be set to correspond to a higher accuracy
temperature estimate; the goal here is simply to compare how well
observations with varying wavelengths can reproduce source
temperatures to a chosen level of accuracy.  We note that at the
-3$\sigma$ level, the difference of $|T_{est} - T_0|$ is always less
than that difference at the +3$\sigma$ level, due to the logarithmic
functional form of the analytic prediction; at lower flux ratios,
corresponding to lower values of the analytic prediction
(eqn. [\ref{ratio2}]), the derived temperature is less sensitive to
uncertainties in the fluxes (see Figs. \ref{lookupval} - \ref{hist1}).

Figure \ref{Tthfig} shows the threshold temperatures, for given
wavelength ratios $\lambda_3/\lambda_2$ and $\lambda_2/\lambda_1$,
where $\lambda_1 < \lambda_2 < \lambda_3$.\footnote{Table \ref{l3tab}
in the Appendix explicitly lists the threshold temperatures $T_{th}$
for various sets of wavelengths, along with the wavelength ratios.
Our choice of the three wavelengths span the range 70 - 3000 $\mu$m.
These particular wavelengths are shown because many of them are
included in the wavebands of the upcoming {\it Herschel} and {\it
Planck} missions.}  For simplicity, we assume a 10\% noise level in
the fluxes at all three wavelengths.  A clear trend is immediately
apparent in Figure \ref{Tthfig}.  The highest values of $T_{th}$ occur
when the two ratios $\lambda_3/\lambda_2$ and $\lambda_2/\lambda_1$
are simultaneously large.  However, for a given ratio
$\lambda_3/\lambda_2$ or $\lambda_2/\lambda_1$ there is a limit to how
large the other ratio can be, beyond which $T_{th}$ decreases.  From
Table \ref{l3tab}, the highest $T_{th}$ is achieved when $\lambda_1 =
70\, \mu$m and $\lambda_3 > 2000\, \mu$m.  At those wavelengths, both
the Wien and R-J regimes of the SED for sources with $T$\aplt\ 75 K
are sampled.

Given two wavelengths that sample the Wien and R-J limits, the maximum
temperature that can be reliably found is set by the middle wavelength
$\lambda_2$: the maximum temperature is roughly that determined by
Wien's displacement law using $\lambda_2$ as the wavelength
corresponding to the flux peak.  This general trend evidently
dissolves as one or more of the boundary wavelengths $\lambda_1$ or
$\lambda_3$ approaches $\lambda_2$.  Since the wavelength
corresponding to the SED peak is inversely proportional to the
temperature, an increase of all three wavelengths by a constant factor
would result in a decrease in $T_{th}$ by that same factor.  Table
\ref{l3tab} indeed shows such a trend.  In short, the exact value of
$T_{th}$ is dependent on all three wavelengths, as would be expected.

We have shown that given three wavelengths, one could determine a
threshold temperature above which the ratio method will not be able to
accurately derive the source temperature (to some chosen level of
accuracy), due to uncertainties in the observations.  In general,
lower temperatures regions, such as cold, dense cores, can be more
accurately measured through this method.  In principle one could also
determine which three wavelengths, and the associated uncertainties,
are optimal to estimate a given temperature.  However, such an
analysis would not be as practical, since observers do not have an
arbitrary choice as to what wavelengths they can observe, nor to any
desired level of accuracy.  After a brief discussion on estimating
$\beta$, we will show that fitting the SED directly to estimate the
temperature is much more accurate than the ratio method, regardless of
what three wavelengths are observed.

\subsection{Estimating $\beta$ \label{betasec}}

In the ratio method, once a temperature estimate is obtained, it may
be used in equation (\ref{betaeqn}) to estimate $\beta$.  Only two
fluxes are required in equation (\ref{betaeqn}).  For isothermal
sources, any two fluxes will give good estimates of $\beta$.  Since
the SED is more sensitive to noise at longer wavelengths, shorter
wavelength fluxes will produce more accurate $\beta$ estimates.
However, determining whether a source is isothermal itself is not
trivial; as discussed in $\S$\ref{obimpsec}, with numerous fluxes this
can be accomplished by systematically excluding short wavelength
fluxes in a direct SED fit.  When a source is not isothermal, the
resulting temperature estimates from short wavelength fluxes will be
highly inaccurate.  The use of these incorrect temperatures in
equation (\ref{betaeqn}) will also produce incorrect $\beta$
estimates.  Thus, an accurate estimate of $T$ is required before
accurately estimating $\beta$ through equation (\ref{betaeqn}).

Uncertainties in observations in the R-J tail will result in highly
inaccurate $\beta$ estimates using the flux ratio method.  For
example, consider an isothermal source that is observed at three
wavelengths, 450, 850, and 1200 \micron.  Fluxes that are inaccurate
by a mere 3\% can produce $\beta$ estimates that deviate from the
actual value by 25\%.  For the simple 2 component source considered in
$\S$\ref{2comsec} (and $\S$\ref{2wave}), and with flux uncertainties
of only 5\%, the estimated $\beta$ can be inaccurate by $\sim$50\%.
Thus, the flux ratio method gives highly uncertain estimates of
$\beta$.

\section{Comparison of Fitting and Flux Ratio Methods involving Three Fluxes \label{comparesec}}

In this section, using simulated observations of a dense core with
temperature and density gradients, we compare the temperature estimate
from direct SED fit to that derived from a flux ratio.  Fluxes are
only ``observed'' at three wavelengths, and include uncertainty due to
noise, requiring that one of the three parameters in the fit ($T$,
$\beta$, or $N\kappa_0$) is held fixed.

One approach to carry out the comparison would be to construct SEDs
throughout the volume of the core, using equations
(\ref{fd})-(\ref{opac}), assuming a fixed value for $\beta$.  We could
then integrate along all lines of sight to obtain the emergent
intensity at three wavelengths, and then carry out the fitting, as we
did in $\S$\ref{fitsec}, or use the flux ratio method to estimate the
temperature.

An alternative approach, which we choose to use here, is to utilize a
radiative transfer (RT) code.  With an RT simulation, the dust
properties can be set using real dust optical constants.  We thus use
the radiative transfer code {\tt MOCASSIN} \citep{Ercolanoetal05}.
{\tt MOCASSIN} is a 3D code that uses a Monte Carlo approach to the
transfer of radiation through an arbitrary medium
\citep{Ercolanoetal03,Ercolanoetal08}.

Our spherically symmetric model core emits radiation according to the
emissivity properties of the dust.  The physical properties of the
core, such as the density and temperature profiles, as well as the
dust properties, are held constant.  Since we are only interested in
the form the observed SED, and not the detailed distribution, the
computational grid only has 16 $\times$ 16 $\times$ 16 zones, which
follows the propagation of radiation through one-quarter of a
spherical model core, with appropriate boundary conditions.  The
overall dimensions of the grid is $\sim$0.1 pc$^3$; the resolution of
an individual zone is comparable to the resolution of the TMC-1C maps
presented by \citet{SchneeGoodman05} and \citet{Schneeetal07}.  As
radiation (or energy packets) from all spatial locations traverses
through the cloud, it is absorbed and re-emitted by the dust.
However, we maintain the original temperature of the dust throughout
the simulation.  In this sense, the temperature of the core is set by
some external source, such as an ambient interstellar radiation field,
which is not explicitly included in our model.  Radiation that emerges
out of the core contributes to the ``observed'' flux.  The emergent
flux will thus be proportional to the density along the line of sight;
the resulting observed 2D map of the core will scale with the 3D
density integrated over the line of sight.  We can then apply the flux
ratio and fitting methods to estimate the temperature at each location
on the 2D map.

We construct a core with a Bonnor-Ebert like density profile, where
the density is constant in the central regions, but then drops off as
the square of the radius.  The temperature of this core is also
constant in the central regions, but then increases logarithmically
with radius.  The temperature of the core varies from $\sim$7.5 K at
the inner regions, to \apgt\ 12 K at the edge of the core.  Such
values span temperatures near the threshold temperature of the flux
ratio method using 450, 850, and 1200 $\mu$m fluxes, as well as
temperatures well into the regime where the method becomes extremely
sensitive to noise.

Figure \ref{varcore} shows the core temperatures, both the actual
temperature and the recovered ones.  The ratio method produces a
spread of temperatures at any given radius, due to noise introduced by
the finite number of photons tracked in the simulation.  A similar
simulation of a core with constant density and constant temperature
produces normally distributed fluxes with $\sigma \approx$ 10\%.
Despite the scatter in the observed fluxes, a general trend of
decreasing temperature with decreasing radius is apparent.  Towards
the innermost regions, the flux ratio method overestimates the
temperature more so than in the outer regions.  This occurs because in
the inner regions of the 2D observed map there is still a contribution
to the flux from the warmer dust at larger radii, from matter lying
above the colder central regions.  At large radii, line of sight
variations in temperature and density is minimized, so the mean of the
flux ratio estimated temperatures corresponds well with the actual
temperatures.  Using the estimated temperatures from the flux ratio
method in equation \ref{betaeqn} gives $\beta = 2.0 \pm 0.1$.  This is
in good agreement with our choice of silicate grains in the
simulation, which should have $\beta$=2 \citep{Draine&Lee84}.

In performing a fit to only three fluxes at each location, one of the
free parameters must be held fixed.  Since the flux ratio method
produces a $\beta$ estimate that is well constrained, even though $T$
shows relatively higher levels of scatter, we hold $\beta$ fixed at
that value.  As shown in Figure \ref{varcore}, the temperatures
obtained through such a fit are also overestimated at small radii.
But, the scatter in the estimated temperatures is much smaller using a
constrained SED fit (i.e. $\beta$ fixed) compared with the flux ratio
method.

Also plotted on Figure \ref{varcore} is the true column temperature
(i.e. density weighted temperature).  Each location (or grid zone) in
the 2D map corresponds to a line of sight through the 3D core.  We
integrate the temperature along the line of sight at each location in
the 2D map, weighted by the 3D densities used to construct the core.
Evidently, the fit temperatures coincide remarkably well with the
column temperature.  The range in temperatures from the flux ratio
method is also spread more evenly about the column temperature than
the actual 3D temperature.  At the innermost regions, there remains a
slight offset.  In general, though, the temperature estimates are
certainly more representative of the ``column temperature'' than the
true temperature of the core.  This is not too surprising, because the
observed flux indeed encodes information about all matter along a line
of sight.

Had we assumed a different (and thus incorrect) value of $\beta$ that
was held fixed in the fit, the best fit temperature would be
systematically offset from the column temperature.  For a 10 K
isothermal source, fixing $\beta$ at 1.7 and 2.3 would produce fit
temperatures of $\sim$ 12 and $\sim$ 9 K, respectively, using fluxes
at 450, 850, and 1200 \micron.  The results of our simple tests
suggest that though the flux ratio method can give highly uncertain
temperature estimates from three fluxes, due to line of sight
temperature variations, those temperatures can still provide a decent
first estimate for the (mean) spectral index (as found by
\citet{Schneeetal07} for isothermal sources).  This estimate for
$\beta$ can then be held fixed in a fit which will recover
temperatures with less scatter, and in close agreement with the column
temperature.

\section{Discussion \label{discsec}}

A flux ratio method to estimate the temperatures can also be
constructed with four wavelengths.  The form of the analytical
prediction for four wavelengths is similar to expression
(\ref{ratio2}).  We have also performed Monte Carlo simulations to
test whether a flux ratio method using four wavelengths produces more
accurate estimates of the source temperature, compared with a flux
ratio method involving only three wavelengths.  The general trends we
found for three wavelengths remains: lower wavelength observations
produce less scatter in the estimated temperatures, and that the
fluxes should sample different regions of the spectrum to obtain
decent temperature estimates.  However, the maximum temperature at
which the turn over occurs for any set of 4 wavelengths listed in
Table \ref{l3tab} is $\sim$ 65 K, for our definition of $T_{th}$ in
$\S$\ref{noisesec}.

When fluxes at four wavelengths are available, however, a direct fit
may be employed to estimate $\beta$ (and the absolute scaling
$N\kappa_0$) along with the temperature.  Since determining $\beta$
through the the flux ratio (eqn. [\ref{betaeqn}]) could give
contradictory estimates depending on which the fluxes are used, a
direct SED fit is preferable.  We did not find any advantage of using
the flux ratio method to determine the temperature using four
wavelengths compared with a direct fit of a modified blackbody SED.

In all of our tests, we have only considered sources with constant
spectral indices.  A line of sight may also have a variations in
$\beta$, and would further complicate the estimation of dust
temperature.  It may be reasonable to assume that the dust emissivity
is constant within a core, where temperatures only vary by $\sim$ 10
K.  But for lines of sight extending through a wider range in density
and temperature, such as the lower density, warmer gas surrounding
sites of recent star formation, assigning a single value for the
spectral index may lead to errors in determining the temperature.  A
thorough investigation of spectral index variations over a range of
environments would be required to quantify its effect on the emergent
SED.

\section{Summary \label{sumsec}}

We have investigated the effect of noise and line of sight temperature
variations on two common methods used to estimate the dust temperature
and spectral index of cold star forming cores using continuum
observations.  One method is a direct fit to a modified blackbody
spectrum.  The second method involves the use of flux ratios.  

We demonstrate that employing an isothermal modified blackbody
equation (eqn. [\ref{fd}]-[\ref{opac}]) may lead to highly inaccurate
dust temperature and spectral index estimates.  Least squares SED fits
to fluxes in the R-J regime, as opposed to the Wien regime, may
provide accurate spectral index and density weighted temperature, or
column temperature, estimates.  For conditions typical of starless
cores, fluxes in the R-J regime have wavelengths \apgt 600 \micron.
However, the fits to fluxes in the R-J regime are rather sensitive to
observational uncertainties, such as noise.

The flux ratio method may also provide inaccurate parameter estimates
due to line of sight temperature variations, and is also very
sensitive to noise.  In a comparison of the flux ratio and least
squares fitting methods when only three fluxes are available, we find
that a direct fit with the spectral index held fixed provides more
accurate estimates of the column temperature.  The flux ratio method
can be initially used to estimate the value of the spectral index to
be held fixed for a least-squares SED fit.

We summarize our main findings in more detail here:

1) Line of sight temperature variations can lead to inaccurate
temperature and spectral index estimates when fitting a
power-law-modified blackbody SED to observed fluxes.  Near the SED
peak of sources with temperature variations, the spectrum is poorly
fit by an isothermal spectrum.  For longer wavelength observations in
the Rayleigh-Jeans regime of the spectrum, and with minimal
observational uncertainties, a fit can accurately recover the spectral
index (if it is constant), and provides a good estimate of the upper
limit of the column temperature.  However, at these long
(Rayleigh-Jeans) wavelengths a fit is extremely sensitive to noise.

2) Short wavelength observations ($\lambda$ \aplt\ 600 \micron) are
still useful, for they can indicate whether an observed source
contains temperature variations.  For starless-core-like sources with
temperature variations, the resulting fit $T$ decreases and fit
$\beta$ increases when systematically excluding short wavelength
fluxes from the fit.  Published data of sources in Taurus and Orion by
\citet{Stepniketal03} and \citet{Dupacetal01}, respectively, show
these apparent trends, but an isothermal description with no
systematic variations in $\beta$ still cannot be strictly ruled out,
due to the uncertainties.  Observed fluxes by \citet{Kirketal07} of
B68, though, produce lower fit temperatures and higher fit spectral
indices when short wavelength fluxes are omitted, strongly suggestive
of dust temperature variations along the line-of-sight.  We estimate
an upper limit of 10.8 $\pm$ 0.1 K for the temperature of the coldest
region within B68; and if the spectral index is constant throughout
the core, then we estimate $\beta \geq$ 2.4.

3) SED fits to fluxes in the Rayleigh-Jeans regime are very sensitive
to noise, even for isothermal sources.  The fits may produce a
spurious inverse $T$-$\beta$ relationship, similar to the trend
discussed by \citet{Dupacetal03}.  SED fits may be more accurate when
fluxes with wavelengths that span the SED peak are available, compared
with fits to fluxes solely in the Rayleigh-Jeans regime.  However,
fits to fluxes near the SED peak would be inaccurate if the source
contains line of sight temperature variations.  In general, for
objects that are cool pockets in higher density regions, such as
starless cores, SED fits that produce higher temperatures also
(artificially) give lower spectral indices.

4) We find that, due to noise uncertainties in any observation, the
flux ratio method is most accurate for emission originating from cold
isothermal regions.  For a source with a constant temperature, there
may be a range in the estimated temperature, due to the uncertainties
in the observations.  At low temperatures, this spread is small; at
higher temperatures the spread can be rather large, rendering the
temperature estimate highly inaccurate.  The precise temperature, or
threshold temperature, for which the method shifts from relatively
accurate to inaccurate is dependent on the observed wavelengths (as
well as the desired level of accuracy, see Fig. \ref{Tthfig} and Table
\ref{l3tab}).  For example, for fluxes at 450, 850, and 1200 \micron,
the flux ratio method can provide accurate temperature estimates only
for sources with $T$ \aplt 7 - 10 K (Fig. \ref{lookupval}).

5) Using Monte Carlo simulations, we quantified the dependence of the
turn over temperature on the set of observed wavelengths.  Ideally, as
one might intuitively expect, two of the three wavelengths should
sample the Wien and the Rayleigh-Jeans regime of the SED, with the
final wavelength lying at intermediate values.  Further, a greater
separation between the wavelengths results in more accurate
temperature estimates.  In general, higher temperatures can be more
accurately measured when the observations include short wavelength
far-infrared observations $\lambda$ \aplt 100 $\mu$m; at short
wavelengths, however, there may be a contribution from stars and
transiently heated very small grains to the observed flux.

6) A reasonably accurate estimate of $\beta$ can be obtained from the
mean of the estimates derived from the flux ratio method involving
three fluxes.  For fluxes at 450, 850, and 1200 \micron, with
$\sim$10\% uncertainties, $\beta$ can be estimated to within 5\% of
the true source value.  This value can then be held fixed in a
constrained SED fit to the three fluxes to estimate the temperature
with less scatter than an estimate from the flux ratio method
(Fig. \ref{varcore}).  With four or more observations $\beta$ may be
one of the free parameters in the fit (and, of course, the fit is
better constrained).

7) The temperatures estimated through the SED fit and ratio methods,
however, cannot be used to assign the absolute temperature to a given
3D location in a cold core.  The projected SED contains information
from all emitting matter along any line of sight.  The measured
temperature is more representative of the column temperature.  In this
regard, the estimated temperatures provide an upper limit for the
coldest temperature along the line of sight.

\appendix
\section{Appendix}
\begin{figure}
\plotone{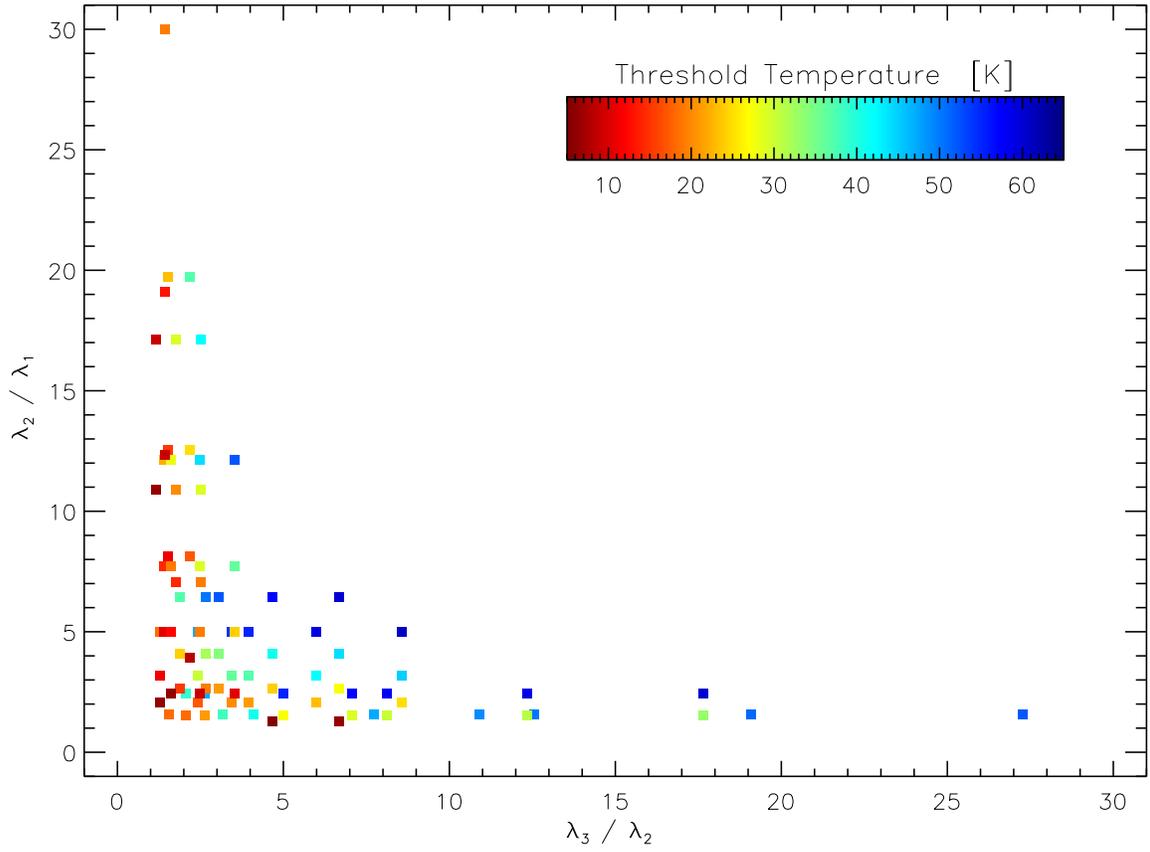}
\caption{Threshold temperature $T_{th}$ (see $\S$\ref{noisesec})
from the flux ratio method, for given ratios $\lambda_3/\lambda_2$ and
$\lambda_2/\lambda_1$, where $\lambda_1 < \lambda_2 < \lambda_3$. }
\label{Tthfig}
\end{figure}

\begin{figure}
\plotone{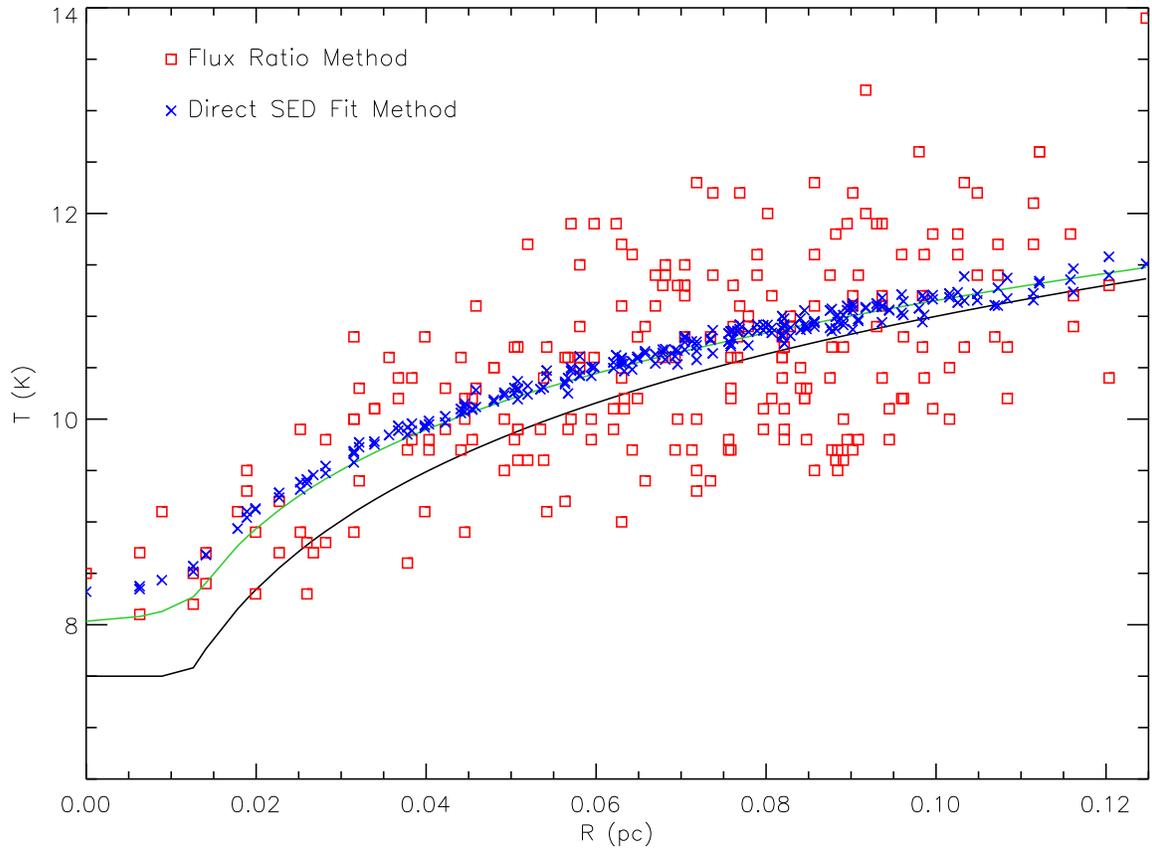}
\caption{Comparison of flux ratio and SED fitting methods to estimate
fluxes.  Solid black line shows actual (3D) temperature profile of
core.  The red squares show temperatures derived using flux ratio
method.  Blue crosses show best fit SED temperatures assuming a fixed
value of $\beta$ = 2.0.  The green line shows the integrated
temperature along the line of sight, weighted by the density, or
``column temperature.''}
\label{varcore}
\end{figure}

Table \ref{l3tab} shows the threshold temperatures $T_{th}$ given
fluxes at three wavelengths (see $\S$ \ref{noisesec}).  The ratios
$\lambda_3/\lambda_2$ and $\lambda_2/\lambda_1$ are also provided.
The table explicitly shows the quantities used to produce Figure
\ref{Tthfig}.

\acknowledgements 

We thank P. Myers, D. Johnstone, J. Foster, J. Pineda, E. Rosolowsky,
and S. Chakrabarti for useful discussions.  We also thank N. Wright
for help in executing MOCASSIN.  In our analysis, we have made
extensive use of NEMO software \citep{Teuben95}.  S. S. acknowledges
support from the Owens Valley Radio Observatory, which is supported by
the National Science Foundation through grant AST 05-40399.  R.S.,
J. K., and A. G. acknowledge support from the Harvard Initiative in
Innovative Computing, which hosts the Star-Formation Taste Tests
Community at which further details on these results can be found and
discussed (see http://www.cfa.harvard.edu/$\sim$agoodman/tastetests).

\bibliography{obsref}

\begin{deluxetable}{cccccccccc}
  \tablewidth{0pt} \tablecaption{Overview of $\S$3 - $\S$6}
\tablehead{ 
\colhead{Scenario} & \colhead{$\S$3.1} & \colhead{$\S$3.2} & \colhead{$\S$3.3} & \colhead{$\S$3.4} & \colhead{$\S$4.1} & \colhead{$\S$4.2} & \colhead{$\S$4.3} & \colhead{$\S$5} & \colhead{$\S$6}}
\startdata
Least Squares Fitting & \checkmark & \checkmark & \checkmark & App\tablenotemark{a} & & & & \checkmark &  \\
Flux Ratio Method & & & & & \checkmark & \checkmark & \checkmark & \checkmark & \\
\hline
$T$ constant & & \checkmark & & & \checkmark & \checkmark & & & \\
Two $T$ Medium & \checkmark & & \checkmark & & \checkmark & & & & \\
Gradient in $T$ & \checkmark & & & & & & & \checkmark & \\
\hline
$\beta$ constant & \checkmark & \checkmark & \checkmark & App\tablenotemark{a} & Fixed\tablenotemark{b} & N/A\tablenotemark{c} & Derived\tablenotemark{d} & \checkmark & \\
$\beta$ variable & & & & & & & & & \checkmark \\
\hline
Two Fluxes & & & & & \checkmark & & & & \\
Three Fluxes & & & & & & \checkmark & & \checkmark & \\
$\#$ Fluxes $>$ 3 & & & \checkmark & App\tablenotemark{a} & & & & & \checkmark \\
$\#$ Fluxes $\gg$ 3 & \checkmark & \checkmark & & & & & & & \\
\hline
Without Noise & \checkmark & & \checkmark & & \checkmark & & & &  \\
With Noise & & \checkmark & & App\tablenotemark{a} & \checkmark & \checkmark & & \checkmark & \\
\enddata
{\singlespace 
\tablenotetext{a}{Application: SED fits to fluxes from published observations}
\tablenotetext{b}{Choice of $\beta$ required in method}
\tablenotetext{c}{$\beta$ is not required or recovered from method}
\tablenotetext{d}{$\beta$ can derived from flux ratio method involving 2 fluxes}
}
\label{overv}
\end{deluxetable}

\begin{deluxetable}{ccccc}
  \tablewidth{0pt} \tablecaption{Temperature Estimates from 2-Flux Ratio Method}
\tablehead{ 
\colhead{Observed $\lambda_1$, $\lambda_2$} & \colhead{Assumed $\beta$\tablenotemark{a}} & \colhead{$\sigma_{obs}$} & \colhead{Source $T_{col}$} & \colhead{Derived $T$} \\
\colhead{(\micron)} & \colhead{} & \colhead{(\%)} & \colhead{(K)} & \colhead{(K)}}
\startdata
Isothermal: \\
450,850 & 1.9 & 0 & 10 & 10.5 \\
450,850 & 2.0 & 0 & 10 & 10.0 \\
450,850 & 2.1 & 0 & 10 & 9.5 \\
450,850 & 1.9 & 10 & 10 & 10.6 $\pm$ 1.3 \\
450,850 & 2.0 & 10 & 10 & 10.0 $\pm$ 1.2 \\
450,850 & 2.1 & 10 & 10 & 9.6 $\pm$ 1.0 \\
\\
1200,2100 & 1.9 & 0 & 10 & 12.1 \\
1200,2100 & 2.0 & 0 & 10 & 10.0 \\
1200,2100 & 2.1 & 0 & 10 & 8.6 \\
1200,2100 & 1.9 & 10 & 10 & 14.5 $\pm$ 12.6 \\
1200,2100 & 2.0 & 10 & 10 & 12.1 $\pm$ 9.4 \\
1200,2100 & 2.1 & 10 & 10 & 10.2  $\pm$ 7.0 \\
\\
\hline
2-Component Source (2COMPd): \\
450,850 & 1.9 & 0 & 10.9 & 12.5 \\
450,850 & 2.0 & 0 & 10.9 & 11.7 \\
450,850 & 2.1 & 0 & 10.9 & 11.0 \\
450,850 & 1.9 & 10 & 10.9 & 12.7 $\pm$ 2.0 \\
450,850 & 2.0 & 10 & 10.9 & 11.8 $\pm$ 1.7 \\
450,850 & 2.1 & 10 & 10.9 & 11.1 $\pm$ 1.5 \\
\\
1200,2100 & 1.9 & 0 & 10.9 & 13.8 \\
1200,2100 & 2.0 & 0 & 10.9 & 11.1 \\
1200,2100 & 2.1 & 0 & 10.9 & 9.3 \\
1200,2100 & 1.9 & 10 & 10.9 & 15.4 $\pm$ 13.4 \\
1200,2100 & 2.0 & 10 & 10.9 & 13.5 $\pm$ 10.9 \\
1200,2100 & 2.1 & 10 & 10.9 & 11.3  $\pm$ 8.2 \\
\enddata
\vspace{-0.5in}
{\singlespace 
\tablenotetext{a}{Model Spectral index $\beta$ = 2.0} 
}
\label{2fluxes}
\end{deluxetable}

\begin{deluxetable}{cccccc} 
  \tablewidth{0pt} \tablecaption{Threshold Temperatures in Ratio
    Method} \tablehead{ \colhead{$\lambda_1$} & \colhead{$\lambda_2$}
    & \colhead{$\lambda_3$} & \colhead{$\lambda_3/\lambda_2$} & \colhead{$\lambda_2/\lambda_1$} & \colhead{$T_{th}$\tablenotemark{a}}}
    \startdata

70 & 110 & 170 & 1.55 & 1.57 & 18 \\ 
70 & 110 & 350 & 3.18 & 1.57 & 38 \\ 
70 & 110 & 450 & 4.09 & 1.57 & 41 \\ 
70 & 110 & 850 & 7.73 & 1.57 & 47 \\ 
70 & 110 & 1200 & 10.91 & 1.57 & 49 \\ 
70 & 110 & 1380 & 12.55 & 1.57 & 50 \\ 
70 & 110 & 2100 & 19.09 & 1.57 & 51 \\ 
70 & 110 & 3000 & 27.27 & 1.57 & 52 \\ 
70 & 170 & 350 & 2.06 & 2.43 & 39 \\ 
70 & 170 & 450 & 2.65 & 2.43 & 47 \\ 
70 & 170 & 850 & 5.00 & 2.43 & 55 \\ 
70 & 170 & 1200 & 7.06 & 2.43 & 57 \\ 
70 & 170 & 1380 & 8.12 & 2.43 & 58 \\ 
70 & 170 & 2100 & 12.35 & 2.43 & 59 \\ 
70 & 170 & 3000 & 17.65 & 2.43 & 60 \\ 
70 & 350 & 450 & 1.29 & 5.00 & 18 \\ 
70 & 350 & 850 & 2.43 & 5.00 & 47 \\ 
70 & 350 & 1200 & 3.43 & 5.00 & 54 \\ 
70 & 350 & 1380 & 3.94 & 5.00 & 55 \\ 
70 & 350 & 2100 & 6.00 & 5.00 & 59 \\ 
70 & 350 & 3000 & 8.57 & 5.00 & 61 \\ 
70 & 450 & 850 & 1.89 & 6.43 & 37 \\ 
70 & 450 & 1200 & 2.67 & 6.43 & 50 \\ 
70 & 450 & 1380 & 3.07 & 6.43 & 52 \\ 
70 & 450 & 2100 & 4.67 & 6.43 & 57 \\ 
70 & 450 & 3000 & 6.67 & 6.43 & 60 \\ 
70 & 850 & 1200 & 1.41 & 12.14 & 22 \\ 
70 & 850 & 1380 & 1.62 & 12.14 & 28 \\ 
70 & 850 & 2100 & 2.47 & 12.14 & 44 \\ 
70 & 850 & 3000 & 3.53 & 12.14 & 52 \\ 
70 & 1200 & 1380 & 1.15 & 17.14 & 9 \\ 
70 & 1200 & 2100 & 1.75 & 17.14 & 29 \\ 
70 & 1200 & 3000 & 2.50 & 17.14 & 42 \\ 
70 & 1380 & 2100 & 1.52 & 19.71 & 23 \\ 
70 & 1380 & 3000 & 2.17 & 19.71 & 37 \\ 
70 & 2100 & 3000 & 1.43 & 30.00 & 19 \\ 
110 & 170 & 350 & 2.06 & 1.55 & 18 \\ 
110 & 170 & 450 & 2.65 & 1.55 & 21 \\ 
110 & 170 & 850 & 5.00 & 1.55 & 27 \\ 
110 & 170 & 1200 & 7.06 & 1.55 & 29 \\ 
110 & 170 & 1380 & 8.12 & 1.55 & 30 \\ 
110 & 170 & 2100 & 12.35 & 1.55 & 31 \\ 
110 & 170 & 3000 & 17.65 & 1.55 & 33 \\ 
110 & 350 & 450 & 1.29 & 3.18 & 11 \\ 
110 & 350 & 850 & 2.43 & 3.18 & 30 \\ 
110 & 350 & 1200 & 3.43 & 3.18 & 36 \\ 
110 & 350 & 1380 & 3.94 & 3.18 & 37 \\ 
110 & 350 & 2100 & 6.00 & 3.18 & 42 \\ 
110 & 350 & 3000 & 8.57 & 3.18 & 45 \\ 
110 & 450 & 850 & 1.89 & 4.09 & 24 \\ 
110 & 450 & 1200 & 2.67 & 4.09 & 32 \\ 
110 & 450 & 1380 & 3.07 & 4.09 & 34 \\ 
110 & 450 & 2100 & 4.67 & 4.09 & 41 \\ 
110 & 450 & 3000 & 6.67 & 4.09 & 44 \\ 
110 & 850 & 1200 & 1.41 & 7.73 & 14 \\ 
110 & 850 & 1380 & 1.62 & 7.73 & 19 \\ 
110 & 850 & 2100 & 2.47 & 7.73 & 29 \\ 
110 & 850 & 3000 & 3.53 & 7.73 & 36 \\ 
110 & 1200 & 1380 & 1.15 & 10.91 & 6 \\ 
110 & 1200 & 2100 & 1.75 & 10.91 & 20 \\ 
110 & 1200 & 3000 & 2.50 & 10.91 & 29 \\ 
110 & 1380 & 2100 & 1.52 & 12.55 & 15 \\ 
110 & 1380 & 3000 & 2.17 & 12.55 & 25 \\ 
110 & 2100 & 3000 & 1.43 & 19.09 & 13 \\ 
170 & 350 & 450 & 1.29 & 2.06 & 6 \\ 
170 & 350 & 850 & 2.43 & 2.06 & 17 \\ 
170 & 350 & 1200 & 3.43 & 2.06 & 20 \\ 
170 & 350 & 1380 & 3.94 & 2.06 & 21 \\ 
170 & 350 & 2100 & 6.00 & 2.06 & 23 \\ 
170 & 350 & 3000 & 8.57 & 2.06 & 25 \\ 
170 & 450 & 850 & 1.89 & 2.65 & 15 \\ 
170 & 450 & 1200 & 2.67 & 2.65 & 19 \\ 
170 & 450 & 1380 & 3.07 & 2.65 & 21 \\ 
170 & 450 & 2100 & 4.67 & 2.65 & 24 \\ 
170 & 450 & 3000 & 6.67 & 2.65 & 27 \\ 
170 & 850 & 1200 & 1.41 & 5.00 & 10 \\ 
170 & 850 & 1380 & 1.62 & 5.00 & 12 \\ 
170 & 850 & 2100 & 2.47 & 5.00 & 19 \\ 
170 & 850 & 3000 & 3.53 & 5.00 & 24 \\ 
170 & 1200 & 2100 & 1.75 & 7.06 & 14 \\ 
170 & 1200 & 3000 & 2.50 & 7.06 & 19 \\ 
170 & 1380 & 2100 & 1.52 & 8.12 & 11 \\ 
170 & 1380 & 3000 & 2.17 & 8.12 & 17 \\ 
170 & 2100 & 3000 & 1.43 & 12.35 & 9 \\ 
350 & 450 & 2100 & 4.67 & 1.29 & 5 \\ 
350 & 450 & 3000 & 6.67 & 1.29 & 6 \\ 
350 & 850 & 1380 & 1.62 & 2.43 & 6 \\ 
350 & 850 & 2100 & 2.47 & 2.43 & 9 \\ 
350 & 850 & 3000 & 3.53 & 2.43 & 10 \\ 
350 & 1380 & 3000 & 2.17 & 3.94 & 8 \\ 

\enddata
\vspace{-0.5in}
{\singlespace \tablenotetext{a}{\footnotesize $T_{th}$ chosen as the
temperature at which $\pm 3\sigma$ is within a factor of 2 of the
source temperature.} }
\label{l3tab}
\end{deluxetable}

\end{document}